\documentclass[sigconf, 9pt]{acmart}

\iftrue
\usepackage{titlesec}
\titlespacing\section{2pt}{5pt plus 1pt minus 1pt}{0pt plus 1pt minus 1pt}
\titlespacing\subsection{2pt}{5pt plus 1pt minus 1pt}{0pt plus 1pt minus 1pt}
\titlespacing\subsubsection{2pt}{5pt plus 1pt minus 1pt}{2pt plus 1pt minus 1pt}
\fi

\iftrue
\usepackage{enumitem}
\setlist{leftmargin=5.08mm}
\fi

\usepackage{amsmath}
\usepackage{mathtools}
\usepackage{array}
\usepackage{amssymb} 
\usepackage{mathrsfs}     
\usepackage{comment}
\usepackage{graphicx}
\usepackage{footnote}
\usepackage[normalem]{ulem}
\usepackage{subfigure}
\usepackage{bm,bbm}
\usepackage{multirow}
\usepackage{threeparttable,booktabs}
\usepackage{blkarray}
\usepackage{tikz}
\usetikzlibrary{positioning,calc,fit,decorations.pathmorphing,shapes.geometric,shapes.gates.logic.US,calc}
\usepackage{balance}
\usepackage{courier}                                       
\usepackage{cleveref}                                      
\usepackage[mathcal]{eucal}
\usepackage{algorithm}
\usepackage{filecontents}                                  
\usepackage{pgfplots}
\usepackage{pgfplotstable}
\usepackage{pgf-pie}
\usepgfplotslibrary{groupplots}
\pgfplotsset{compat=newest}
\usepackage[figuresright]{rotating}
\usepackage{xcolor,colortbl}                               
\usepackage{amsthm}                 

\usepackage{tabularx,  multirow, multicol,  bigstrut}
\usepackage[noend]{algorithmic}

\theoremstyle{plain}

\theoremstyle{definition}

\newtheorem{owndefinition}{Definition}		

\usepackage[skip=1pt]{caption}            
\AtBeginDocument{%
 }

\copyrightyear{2024}
\acmYear{2024}
\setcopyright{acmlicensed}\acmConference[ICCAD '24]{IEEE/ACM International Conference on Computer-Aided Design}{October 27--31, 2024}{New York, NY, USA}
\acmBooktitle{IEEE/ACM International Conference on Computer-Aided Design (ICCAD '24), October 27--31, 2024, New York, NY, USA}
\acmDOI{10.1145/3676536.3676676}
\acmISBN{979-8-4007-1077-3/24/10}

\usepackage{etoolbox}
\makeatletter
\patchcmd{\authornote}{\g@addto@macro\addresses{\@authornotemark}}{}{}{}
\makeatother

\begin{document}
\begin{sloppypar}

\title{Physically Aware Synthesis Revisited: Guiding Technology Mapping with Primitive Logic Gate Placement}

\author{
Hongyang Pan\textsuperscript{1}$^{\ast}$, 
Cunqing Lan\textsuperscript{1}$^{\ast}$, 
Yiting Liu\textsuperscript{1}, 
Zhiang Wang\textsuperscript{2}, 
Li Shang\textsuperscript{1}, 
Xuan Zeng\textsuperscript{1}, 
Fan Yang\textsuperscript{1}, 
Keren Zhu\textsuperscript{1}$^{\dagger}$}
\authornote{Both authors contributed equally to this work.}
\authornote{Keren Zhu is the corresponding author: kerenzhu@ieee.org.}
\affiliation{%
  \institution{\textsuperscript{1}Fudan University, \textsuperscript{2}University of California San Diego}
  \country{}
}









\renewcommand{\shortauthors}{Hongyang Pan, Cunqing Lan et al.}
\renewcommand{\shorttitle}{PigMAP}


\begin{abstract}
    A typical VLSI design flow is divided into separated front-end logic synthesis and back-end physical design~(PD) stages, which often require costly iterations between these stages to achieve design closure.
    Existing approaches face significant challenges, notably in utilizing feedback from physical metrics to better adapt and refine synthesis operations, and in establishing a unified and comprehensive metric. 
    This paper introduces a new \textbf{P}rimitive log\textbf{i}c \textbf{g}ate placement guided technology \textbf{MAP}ping~(PigMAP) framework to address these challenges.
    With approximating technology-independent spatial information, we develop a novel wirelength~(WL) driven mapping algorithm to produce PD-friendly netlists.
    PigMAP is equipped with two schemes: a performance mode that focuses on optimizing the critical path WL to achieve high performance, and a power mode that aims to minimize the total WL, resulting in balanced power and performance outcomes. 
    We evaluate our framework using the EPFL benchmark suites with ASAP7 technology, using the OpenROAD tool for place-and-route.
    Compared with OpenROAD flow scripts, performance mode reduces delay by 14\% while increasing power consumption by only 6\%. Meanwhile, power mode achieves a 3\% improvement in delay and a 9\% reduction in power consumption. 
\end{abstract}

\maketitle

\vspace{-2mm}
\section{INTRODUCTION}
\label{sec:intro}
Recently, the electronic design automation~(EDA) industry has implemented a ``shift-left'' strategy, aiming to accelerate the design cycle by proactively addressing potential issues early in the process~\cite{21IJCA_Bhardwaj_Shiftleft}. 
The growing complexity in achieving timing closure is exacerbated by the mis-correlation between place-and-route and logic synthesis~\cite{11ACM_Li_whatisPD}.
To address this, a synchronized approach known as physically aware synthesis has significantly improved traditional synthesis flows by integrating physical information early to optimize power, performance, and area~(PPA)~\cite{22MDAT_Kahng_MLforEDA}.

Physically aware synthesis typically integrates placement information during the logic synthesis stage, a long-standing objective in the field for decades~\cite{03FPGA_lin_placement,95TCAD_chen_combining,08ICCAD_liu_delay,93_gao_placement}.
This approach is primarily categorized into two classes: gate-level resynthesis~\cite{03FPGA_lin_placement,93_gao_placement,07PI_Alpert_physyntec} and logic-level technology mapping~\cite{23ICCD_Liu_aimap,98DAC_salek_dsm,07TCAD_karandikar_technology,18ALS_Reis_Physical}. 
Historically, efforts to merge placement and logic synthesis began in the 1990s. 
For instance, Lu \emph{et al.}~\cite{98_lu_combining} addressed inaccuracies in delay models used during the mapping phase by incorporating a more accurate post-placement delay model in the logic resynthesis phase. 
In the 2000s, Jiang \emph{et al.}~\cite{99ICCAD_jiang_integrated} introduced additional placement-based constraints in technology mapping to guide the search for favorable solutions. 
More recently, initiatives like those by Liu \emph{et al.}~\cite{11TCAD_Liu_Simultaneous} combined mapping and placement, effectively addressing interconnect impacts by considering potential path delays during initial floorplanning, thus offering a promising solution.
Despite these advancements, recent researches have not been particularly groundbreaking. 
The existing practical applications face three primary limitations:
(1)~Gate-level resynthesis generates physically optimized netlists but offers a limited view of topology, restricting the scope of potential optimizations;
(2)~Logic-level mapping provides a broader perspective but is time-consuming, hindered by inefficient enumeration algorithms and extended placement processes;
and (3)~Simultaneous mapping and placement maintain complete physical information in the mapped netlist but suffer from inefficiencies.

\begin{figure}[t]
    \centering
    \includegraphics[width=0.9\linewidth]{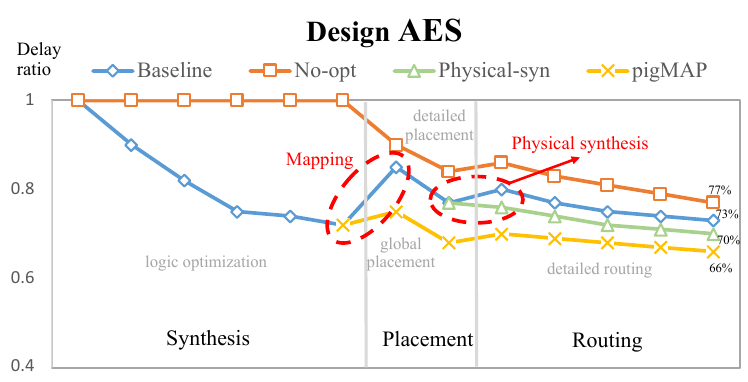}
    \caption{The challenges associated with physically aware synthesis methodology. In design AES~\cite{05IWLS_albrecht_iwls}, the Y-axis shows the ratio of critical path delay, while the X-axis shows the iteration steps, including five rounds of logic optimization, global placement, detailed placement and detailed routing.}
    \label{fig:gap}
    \vspace{-3mm}
\end{figure}

We illustrate the challenges in existing physically aware synthesis methodologies. 
Figure~\ref{fig:gap} displays the delay trends in the synthesis process of AES~\cite{05IWLS_albrecht_iwls} using the end-to-end OpenROAD flow~\cite{24_openroad_script} in the academic \emph{Arizona State Predictive PDK 7nm}~(ASAP7) technology~\cite{16MJ_Clark_asap7}. 
Multiple rounds of technology-independent optimization significantly enhance performance in the front-end phase. 
However, the benefits may diminish in the back-end stage if the mapping cannot effectively manage these optimizations~(blue curve). 
This underscores the importance of effective mapping algorithm. 
Omitting these logic optimizations~(orange curve) does not substantially worsen the final delay, shifting only from 73\% to 77\%. 
This indicates a miscorrelation between critical paths before and after the logic synthesis stage. 
Moreover, physical synthesis~(green curve) can improve post-placement delay, demonstrating that optimizations leveraging placement information can effectively enhance performance. 
Consequently, an effective flow should incorporate an efficient mapping algorithm that utilizes placement-based physical information.

We revisit physically aware synthesis by evaluating the entire design flow and the metrics employed within it. 
Our analysis has identified two major challenges:
\textbf{C1. Maintaining heuristic consistency in the feedback loop of physically aware synthesis}: This challenge ensures that synthesis tools effectively utilize feedback from physical metrics to better adapt and refine mapping operations.
Such adaptability is crucial for making decisions that optimize the physical attributes of the circuit.
\textbf{C2. Formulating new physically relative metrics} to enhance the PPA of circuits: Beyond traditional metrics like delay and area, it is essential to develop new metrics that effectively bridge the gap between synthesis optimizations and their outcomes in the physical design~(PD) phase during the mapping stage.
To address \textbf{C1}, we propose integrating primitive logic gate placement with technology mapping. 
Recent advancements in approximate placement~\cite{22TODAES_Liu_graphplanner} have enabled rapid placement of primitive logic gates.
Combined with technology-independent optimization, this integration allows for a significant ``shift-left'' to implement placement at the logic gate level early in the design flow.
For \textbf{C2}, we introduce a novel metric called ``virtual wirelength'', designed to mitigate the effects of interconnect. 
This metric is computed based on the primitive logic gate placement. 
Inspired by previous work~\cite{11TCAD_Liu_Simultaneous}, we further develop a wirelength-driven technology mapping framework to more precisely generate mapping solution.

In this work, we propose a novel \emph{{\bf P}rimitive log{\bf i}c {\bf g}ate placement guided technology {\bf MAP}ping} (PigMAP) framework.
This framework utilizes the placement of primitive logic gates to rapidly obtain approximate technology-independent spatial information, which is then used to execute wirelength-based mapping. 
To the best of our knowledge, this is the first work that leverages this information to generate PD-friendly netlists.
The key contributions of our work are summarized as follows:
\begin{itemize}[leftmargin=*,itemsep=3pt, topsep=2pt]
    \item {\bf Wirelength-based mapping algorithm}: We present a novel mapping algorithm that improves the quality of results~(QoR) by setting constraints on either the critical path wirelength (PigMAP-Performance) or the total wirelength (PigMAP-Power).
    \item {\bf Primitive logic gate placement}: We pioneer a primitive logic gates placement algorithm that enables fast placement for \emph{And-Inverter Graph}~(AIG). Leveraging techniques from timing prediction algorithms, we estimate potential wirelength between nodes, aiding the mapper in better alignment with subsequent place-and-route steps.
    \item {\bf Physically aware metric}: We propose a new physically relative metric called ``virtual wirelength'' into technology mapping, incorporating placement information to enhance interconnect awareness.
    \item The proposed mapping algorithm is implemented in the open-source logic synthesis tool phyLS.~\footnote{The source code is available in the phyLS GitHub repository~\cite{phyLS}.}
    Experiments using the EPFL benchmark suites show that PigMAP-Performance reduces post-routing delay by 14\% with only a 6\% increase in power consumption, while PigMAP-Power achieves a 3\% improvement in post-routing delay and a 9\% reduction in power consumption.
\end{itemize}

\section{BACKGROUND}
\label{sec:bg}

In this section, we review the background of Boolean networks, technology mapping, recent advancements in physically aware synthesis, and place-and-route processes. Table~\ref{tab:para} summarizes important terms and their respective meanings.

\begin{table}[h]
    \centering
    \fontsize{8}{9}\selectfont
    \caption{Notation and Parameters of PigMAP framework.}  
    \label{tab:para}
    \begin{tabularx}{\hsize}{@{}@{\extracolsep{\fill}}|l|X|@{}}  
    \hline 
    Term & Description \\
    \hline 
    $\mathbf{M}$   &   Mapping solution  \\
    \hline
    $\mathbf{S}$   &   Mapping strategies, given PigMAP-Performance or PigMAP-Power   \\
    \hline
    $\mathbf{N}$        & The computed position of each node   \\    \hline
    $\mathbf{C}$, $\mathbf{C_n}$      & The cuts' set all nodes and cuts' set of the node $n$ \\  \hline
    $\mathbf{P}$, $\mathbf{P_{c}}$     & The pins' set of all cuts, and the pins' set of  of the cut $c$ \\ \hline
    $\mathcal{C}_t$, $\mathcal{C}_{w_t}$, $\mathcal{C}_{w_c}$     & Constraints of timing, global wirelength and critical path wirelength, respectively    \\     \hline  
    $G = (V, E)$   & Graph G with nodes V and edges E \\ \hline
    
    \end{tabularx}   
\end{table}

\subsection{Boolean Network and AIG}
\label{subsec:aig}

A Boolean network can be represented as a directed acyclic graph (DAG) $G=(V,E)$ with a set of nodes $V$ and edges $E$, where $V$ correspond to logic gates (Boolean functions) and $E$ represent the wires connecting these gates.
This DAG configuration can also be referred to as a primitive logic gate network.
Each node $v \in V$ has inputs (outputs), termed fanin (fanout). 
Nodes without incoming edges serve as the primary inputs~(PIs) of the graph.
Similarly, nodes without outgoing edges are the primary outputs~(POs) of the graph.
Nodes with incoming edges implement Boolean functions. 
The level of a node $v$ is defined by counting the nodes along the longest structural path from any PI to $v$, including both endpoints. 
The cut $C$ of node $v$ includes a subset of nodes within the network, referred to as leaves, such that each path from the PIs to $v$ passes through at least one leaf.
Node $v$ is termed the root of the cut $C$.
The cut size is the number of its leaves, and the cut is deemed $k$-feasible if it contains no more than $k$ nodes.

An AIG is a common type of DAGs used in logic manipulations, where each node is a two-input AND gate, and the edges indicate the presence of inverters. Any Boolean network can be converted into an AIG by decomposing the sum of products~(SOPs) of the nodes, transforming the AND and OR gates in SOPs into two-input AND gates and inverters using DeMorgan’s rule.
There are two primary metrics for evaluating an AIG: size and depth. Size refers to the number of nodes~(AND gates) in the graph, while depth is measured by the number of nodes on the longest path from PIs to POs, indicating the highest level within the graph~\cite{06IWLS_Alan_ScalableLS}.

\subsection{Technology Mapping}
\label{subsec:ls}

Technology mapping is the process of translating a Boolean network into a set of primitives defined in a technology library. 
Initially, the Boolean network is represented as a $k$-bounded network, known as the subject graph, which comprises nodes each with maximum fanins of $k$. 
AIGs are commonly used as subject graphs for this purpose. 
Although technology mapping typically employs the state-of-the-art (SOTA) open-source logic synthesis tool ABC~\cite{24_abc}, for effective integration in this work, we utilize Mockturtle~\cite{24_EPFL_mockturtle}, an open-source synthesis implementation toolkit known for its flexible data structures.

The mapping algorithm involves three primary steps:
(1)~Identifying $k$-feasible cuts through a rapid enumeration procedure~\cite{98FPGA_pan_new}; (2)~Associating these cuts with elements from the technology library via Boolean matching~\cite{97TODAES_benini_survey}; and (3)~Creating a graph cover that optimizes a specific cost function while adhering to preset constraints. 
The process is visually represented in Figure~\ref{fig:patmflow} by blue cubes, which includes delay-driven mapping and area-driven mapping.

\subsubsection{Delay-driven mapping. }
\label{subsubsec:ddm}

Delay-driven mapping seeks to optimize the timing performance by identifying the structure with the best timing characteristics, thus minimizing the worst delay of the entire circuit~\cite{04ICCAD_chen_daomap}. 
In topological order, the mapper iteratively visits nodes and traverses every available supergate and cut for each node. 
Gates and cuts that offer the best timing performance are retained as candidates for the mapping solution.
\begin{owndefinition} The delay of a node $n$ (D($n$)) is the maximum delay among all leaves of $n$ plus the delay of the mapped standard cell ($\text{d}(n)$), defined as:
\label{def0}
    \begin{equation}
    \label{eq:delay}
        \text{D}(n)=\max_{i \in leaves(n)}\text{D}(i) + \text{d}(n),
    \end{equation}
    where $\text{d}(n)$ is zero for PIs, and $i$ can be any leaf of $n$.
\end{owndefinition}
After mapping is complete, a fast traversal from POs to PIs is conducted to gather information such as the total area and the worst path's delay of the mapping solution. 
Following the delay mapping, timing constraints are computed by backpropagating the worst delay from POs to PIs.
This backpropagation ensures that timing performance is maintained while optimizing other metrics.

\subsubsection{Area-driven mapping. }
\label{subsubsec:adm}

Area-driven mapping operates on the mapped solution under timing constraints to perform global area optimization.
An area-oriented mapping algorithm that supports multi-output cells is employed to further optimize the area~\cite{06TCAD_manohararajah_heuristics}. 
This algorithm traverses nodes in topological order, selecting matches with the minimum area from each visiting node to PIs, while ensuring that delay constraints are not violated. Each edge and node in the graph is associated with an area flow (AF), which represents an estimate of the mapping solution's area.
\begin{owndefinition} The area flow of a node $n$ (AF($n$)) is the approximate area covering from $n$ to all accessible PIs, defined as:
\label{def1}
    \begin{equation}
    \label{eq:af}
        \text{AF}(n) = A_n + \sum_{i \in leaves(n)} \frac{\text{AF}\left(best\_gate\left(i\right)\right)}{ \text{EstFanouts}\left(best\_gate\left(i\right)\right)},
    \end{equation}
    where the area of a node $A_n$ is zero for PIs, $\text{EstFanouts}$ represents the estimate number of gates driven by $n$ in the current mapping solution. $leaves(n)$ denotes the nodes that paths from PIs to the cut's root $n$ must pass, and $best\_gate(i)$ indicates the optimal gate matched for corresponding leaf $i$.
\end{owndefinition}
This equation considers the area flowing into a node as the total area entering through its input edges. 
The area flowing out of a node includes the area incoming to the node plus an additional component representing the area of the gate currently matched for the node itself. 
Thus, the area flowing out of a node is evenly distributed among its outgoing edges.

\subsection{Physically aware Synthesis}
\label{subsec:pas}

Physically aware synthesis utilizes actual design and physical library information to enhance the synthesis process.  
Placement-based features and their variants commonly serve as constraints in mapping algorithms. 
Additional features include those based on load distribution~\cite{07TCAD_karandikar_technology}, congestion~\cite{02ICCAD_chatterjee_new}, and post-routing~\cite{12DAC_Li_guidingPDclosure,23ICCD_zhu_delay}.
Moreover, graph neural networks based machine learning algorithms are adept at predicting net-length, timing, and congestion via graph attention networks~\cite{19VLSISOC_kirby_congestionnet, 22TCAD_xie_preplacenetlength,23ICCAD_agiza_graphsym}.
Research has also explored logic optimization methods that utilize placement information to enhance performance~\cite{18ALS_Reis_Physical,01ICCAD_gosti_addressing}, as well as the simultaneous mapping and placement to ensure that the mapped netlist preserves complete physical information~\cite{94ITFECCS_togawa_maple,98TCAD_togawa_maple,08ICCAD_liu_delay,97ICCAD_lou_exact,98DAC_salek_dsm}.

\begin{figure}[t]
  \centering  
  \subfigure[Conventional physically aware synthesis methodology]{
    \label{fig:flowps2}
    \includegraphics[width=0.47\linewidth]{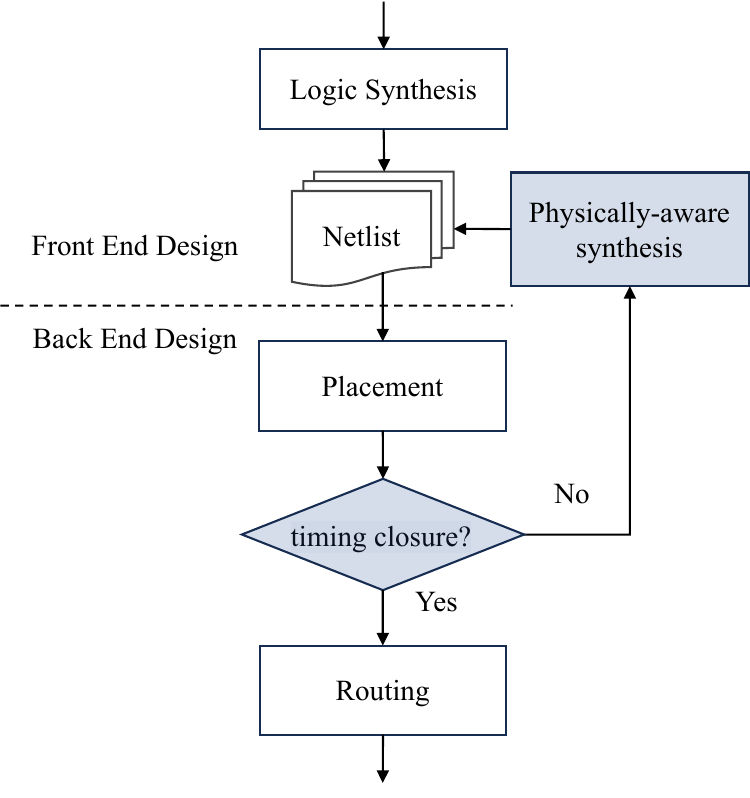}}
  \subfigure[Our proposed flow with placement-guided technology mapping]{
    \label{fig:flowps1}
    \includegraphics[width=0.47\linewidth]{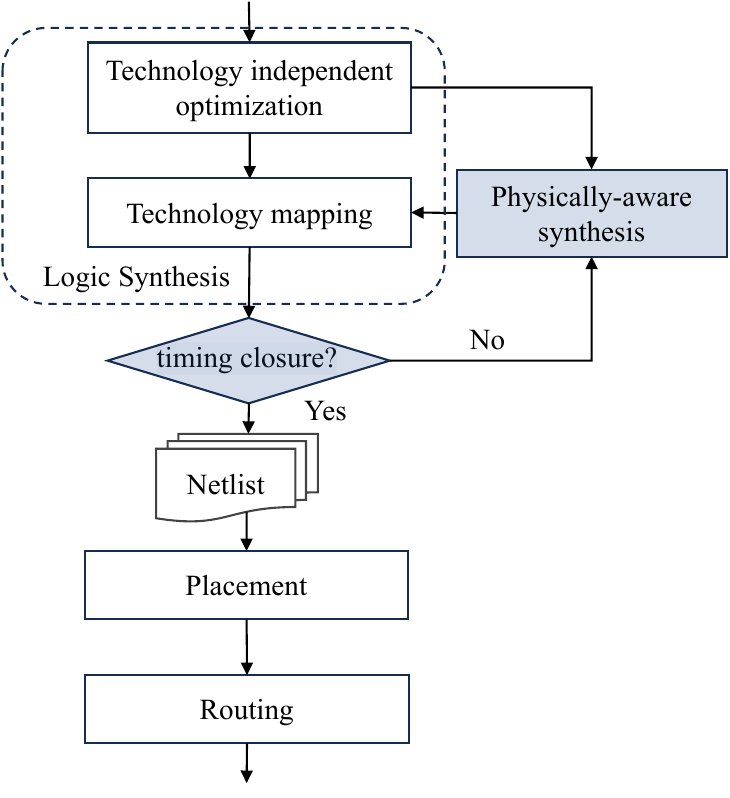}}
  \caption{Different physically aware synthesis flows.}
  \label{fig:flowps}
\end{figure}

To integrate these techniques into the conventional synthesis workflow, additional steps are introduced as depicted in Figure~\ref{fig:flowps2}. 
The process starts with logic synthesis, followed by multiple iterations of physically aware synthesis on the post-placement netlist until timing closure is achieved. 
However, our approach, illustrated in Figure~\ref{fig:flowps1}, advances this concept further ``left''. 
It begins with technology-independent optimization, followed by physically aware synthesis during the technology mapping. 
The mapping process continues until timing closure. 
This methodology establishes a foundational flow for physically aware synthesis, instrumental in creating diverse physical flows to achieve optimal timing results.

\begin{figure*}[t]
    \centering
    \includegraphics[width=0.9\linewidth]{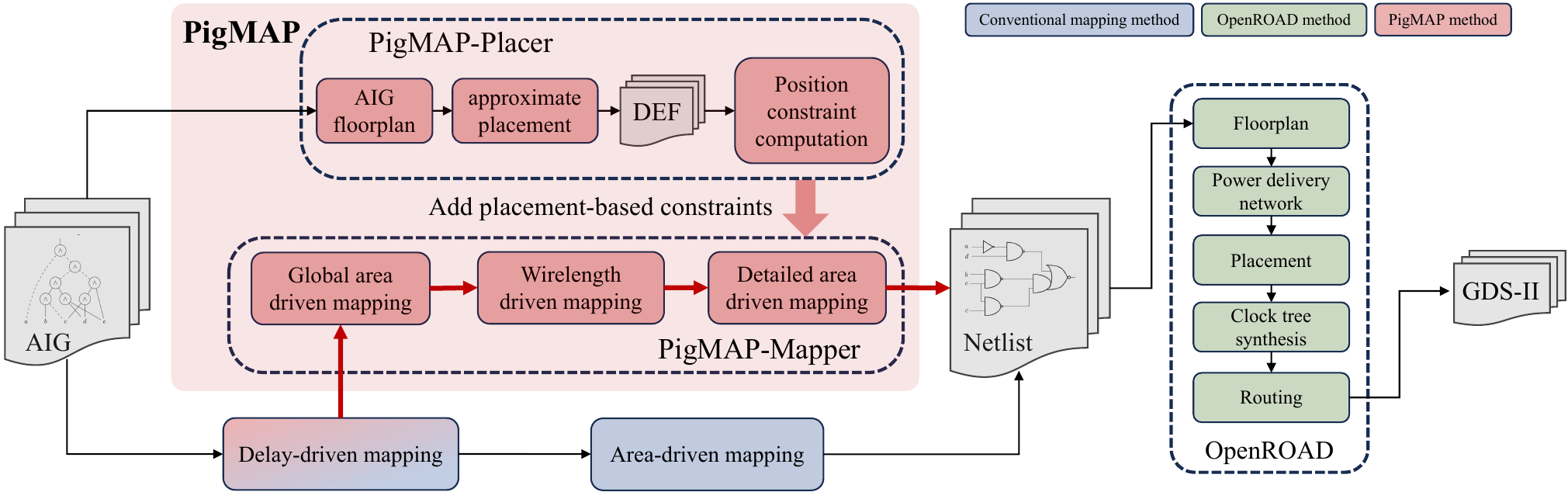}
    \caption{The PigMAP framework of ASIC design.}
    \label{fig:patmflow}
    \vspace{-3mm}
\end{figure*}

\subsection{Place-and-route}
\label{subsec:pr}

A placement instance can be formulated as a graph with a set of objects (e.g. IOs, macros, and standard cells) and edges. The main objective of placement is to find a solution with minimized total wirelength subject to density constraints (i.e., the density does not exceed a predetermined density).
The central challenge is the placement of logic gates or cells on a 2D plane, which requires balancing multiple objectives within various constraints. 
This process is analogous to partitioning in the mapping phase, where logic is strategically segmented. 
Utilizing placement information as constraints can significantly aid the mapper in selecting optimal partitions, thereby enhancing the overall effectiveness of the physically aware synthesis.
Routing begins with processing the technology files and the output from the placement stage, followed by detailed routing for both signal and clock nets, guided by a global routing solution presented in the form of a routing guide.
In this paper, we utilize OpenROAD~\cite{19PGMACTC_Ajayi_openroad} as our layout generation tool chain.

\vspace{-3mm}
\section{OUR APPROACH}
\label{sec:method}

In this section, we introduce our PigMAP framework.  
PigMAP uses wirelength constraints to guide the mapping process, aiming at optimizing critical path delay and power consumption.
Figure~\ref{fig:patmflow} and Algorithm~\ref{alg:PigMAP} present the overall framework of our PigMAP. 
As highlighted by red cubes in Figure~\ref{fig:patmflow}, PigMAP consists of two main components:

\begin{itemize}[leftmargin=*,itemsep=3pt, topsep=2pt]
    \item \textbf{PigMAP-Placer} begins by conducting AIG-based floorplanning, followed by approximate placement to establish the preliminary location of each AIG node (Line 1). 
    
    \item \textbf{PigMAP-Mapper} initializes the data structures for mapping, computing and matching cuts, and computing the inter-node distances of the AIG nodes (Lines 2-6).
    After the initial setup, the PigMAP-Mapper utilizes delay-driven and global area-driven approaches~\cite{23ICCAD_calvino_technology} to map the input AIG. 
    After each mapping iteration, delay constraints are updated to maintain timing estimation (Lines 7-10). 
    Operating under the timing constraints, the PigMAP-Mapper further performs two optimization iterations of a wirelength-driven mapping, specifically aiming at minimizing the wirelength of the final mapped netlist. 
    During the wirelength-driven mapping phase, two alternative refinement strategies are available: performance-driven mapping (\textbf{PigMAP-Per\-for\-mance}), which focuses on the critical path wirelength (Definition~\ref{def2}), and power-driven mapping (\textbf{PigMAP-Power}), which targets reduction in overall global wirelength (Definition~\ref{def3}) (Lines 11-16).
    Following these, the PigMAP-Mapper executes two rounds of heuristic detailed area optimization~\cite{23ICCAD_calvino_technology} to precisely minimize the area at each node under the wirelength constraints, further reducing the area of the final mapped netlist (Line 17).
    With resultant mapping solution given by mapping, we cover the AIG from POs to PIs through visiting leaves and combining the ``best supergate'' saved in each visited node, generating the final netlist (Line 18).
\end{itemize}

\begin{algorithm}[t]
    \footnotesize
    \caption{The PigMAP framework}
    \label{alg:PigMAP}
    \begin{algorithmic}[1]
    \REQUIRE Logic network AIG $\mathbf{A}$ and mapping strategy $\mathbf{S}$
    \ENSURE Gate-level netlist
    \STATE All node position $\mathbf{N} \leftarrow$ \texttt{Placer}($\mathbf{A}$); {\hfill//Algorithm~\ref{alg:aigplace}}
    \STATE Mapping solution $\mathbf{M}$;
    \STATE Compute and match cuts $\mathbf{C}$, and initialize pins $\mathbf{P}$;
    \FORALL{node $n$ in $\mathbf{A}$}
        \FORALL{cut $c$ in $\mathbf{C_n}$} 
            \STATE $\mathbf{P}_c \leftarrow$ \texttt{searchPin}($c$); 
            {\hfill//Algorithm~\ref{alg:search_pins}}
        \ENDFOR
    \ENDFOR
    \STATE $\mathbf{M} \leftarrow$ \texttt{delayMap}($\mathbf{A}, \mathbf{C}$); {\hfill//Execute mapping}
    \STATE Get global time constraint $\mathcal{C}_{t}$;
    \STATE $\mathbf{M} \leftarrow$ \texttt{globalAreaMap}($\mathbf{A}, \mathbf{C}, \mathcal{C}_{t}$);
    \STATE Update global time constraint $\mathcal{C}_{t}$;
    \IF{$\mathbf{S}$ == PigMAP-Power} 
        \STATE $\mathbf{M} \leftarrow$ \texttt{PigMAP\_Power}($\mathbf{A}, \mathbf{C}, \mathcal{C}_{t}, \mathbf{N}, \mathbf{P}$); {\hfill//Algorithm~\ref{alg:paflow}}
        \STATE Get global wirelength constraint $\mathcal{C}_{w_t}$;
    \ELSE
        \STATE $\mathbf{M} \leftarrow$ \texttt{PigMAP\_Performance}($\mathbf{A}, \mathbf{C}, \mathcal{C}_{t}, \mathbf{N}, \mathbf{P}$); {\hfill//Algorithm~\ref{alg:paflow}}
        \STATE Get critical path wirelength constraint $\mathcal{C}_{w_c}$;
    \ENDIF
    \STATE $\mathbf{M} \leftarrow$ \texttt{detailAreaMap}($\mathbf{A}, \mathbf{C}, \mathcal{C}_{t}, \mathcal{C}_{w_t}, \mathcal{C}_{w_c}$);
    \RETURN \texttt{genNetlist}($\mathbf{M}$);
    \end{algorithmic}
\end{algorithm}

\subsection{PigMAP-Placer: Primitive Logic Gate Placement}
\label{subsec:ap}

To incorporate wirelength as a mapping metric, it is necessary to compute the positions and inter-node distances within the AIG. 
While modern analytical placers view the placement problem as a nonlinear optimization process, often resulting in prolonged iteration time, a fully optimized placement for mapping constraints is not always necessary and might be redundant. 
Instead, our method requires only optimized relative positions between nodes.
To achieve this, we employ a graph convolution-based placer to accelerate the process and efficiently generate optimized placement results. 
More details are available in~\cite{24ICCAD_Liu_gift}.

\begin{algorithm}[b]
    \footnotesize
    \caption{PigMAP-Placer Algorithm}
    \label{alg:aigplace}
    \begin{algorithmic}[1]
    \REQUIRE An AIG $\mathbf{A}$
    \ENSURE Optimized locations of nodes $g'$
    \STATE A circuit netlist modeled by an undirected weighted graph $\leftarrow$ FloorPlan($\mathbf{A}$);
    \STATE Locations of movable cells $g_m \sim N(0,1)$ are randomly placed at the center of the placement region;
    \STATE Fixed cells are set at their predetermined locations;
    \STATE Compute optimized node locations $g'$ using graph-convolution-based placer;
    \RETURN $g'$;
    \end{algorithmic}
\end{algorithm}

\begin{figure}[h]
  \centering  
  \subfigbottomskip=5pt 
    \subfigure[AIG network of C17]{
    \label{fig:aig1}
    \includegraphics[width=0.4\linewidth]{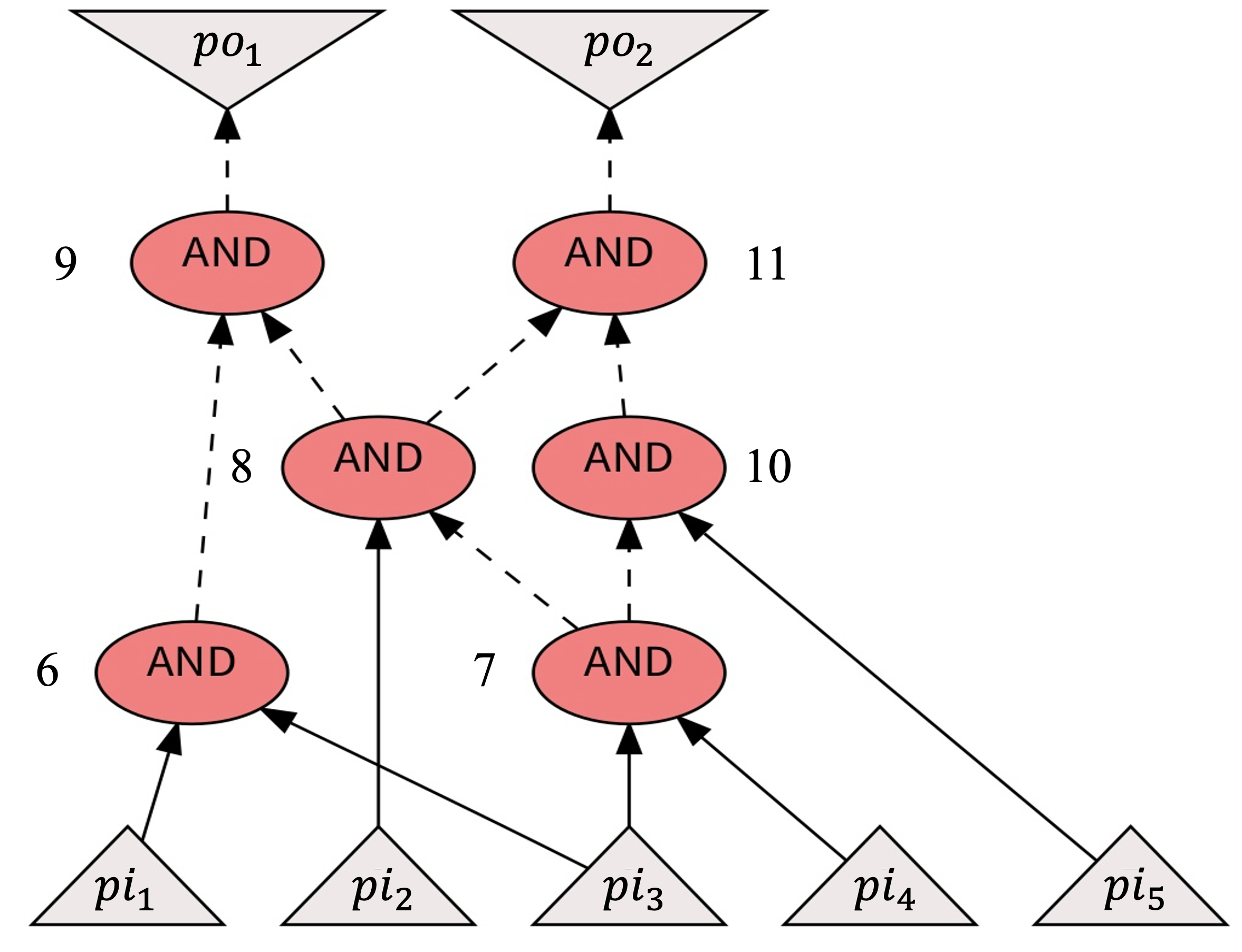}}
    \subfigure[final cell locations]{
    \label{fig:aig6}
    \includegraphics[width=0.42\linewidth]{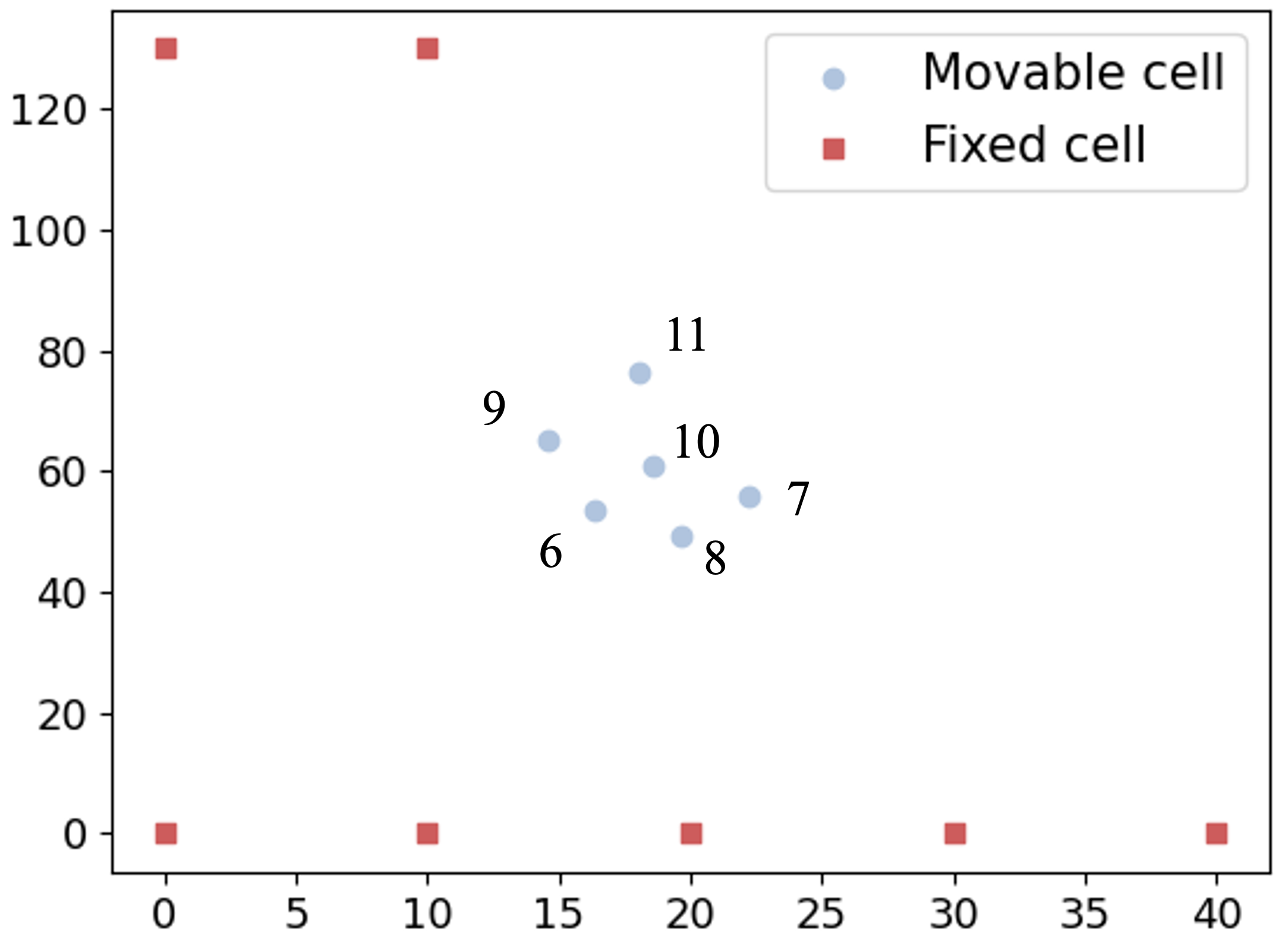}}
  \caption{An example for AIG placement of {\em C17}.}
  \label{fig:aig}
  \vspace{-2mm}
\end{figure}

The process of the PigMAP-Placer is detailed in Algorithm~\ref{alg:aigplace}. 
The entire process is illustrated with the example shown in Figure~\ref{fig:aig}.
In this example, the AIG $\mathbf{A}$ has five PIs \{$pi_1$, $pi_2$, $pi_3$, $pi_4$, $pi_5$\}, two POs \{$po_1$, $po_2$\}, and six intermediate nodes \{$6, 7, 8, 9, 10, 11$\}. 
Initially, AIG-based placement requires floorplanning. 
The nodes of $\mathbf{A}$ are modeled as a cell.
The core area dynamically scales based on the AIG size, with PIs and POs of AIG designated as I/O and fixed in position. 
Next, the PigMAP-Placer randomly places movable cells following a Gaussian distribution, while fixed cells are set at their predetermined locations (Lines 1-3).
Our approach not only learns the optimized mapping between circuit connectivity and physical wirelength but also decodes and determines the optimized initial physical locations of the nodes (Line 4).
Finally, the optimized locations of nodes is obtained, as shown in Figure~\ref{fig:aig6}.
Within AIG, it might initially appear that node $8$ and node $9$ are relatively close to each other. 
However, after placement, it becomes evident that nodes $6$ and $9$ are actually closer. 
By mapping nodes $6$ and $9$ into one cell, we can effectively reduce the wirelength of the netlist.

\subsection{PigMAP-Mapper: Placement-aware Mapping}
\label{subsec:PigMAP}

In this subsection, we introduce our placement-aware mapping algorithm with wirelength constraints.
The PigMAP-Mapper consists of two main steps: preparation and executing.

\begin{figure}[h]
  \centering  
  \subfigbottomskip=5pt 
    \subfigure[pins \{$7,8$\} of cut \{$2,3,4$\}]{
    \label{fig:pin1}
    \includegraphics[width=0.4\linewidth]{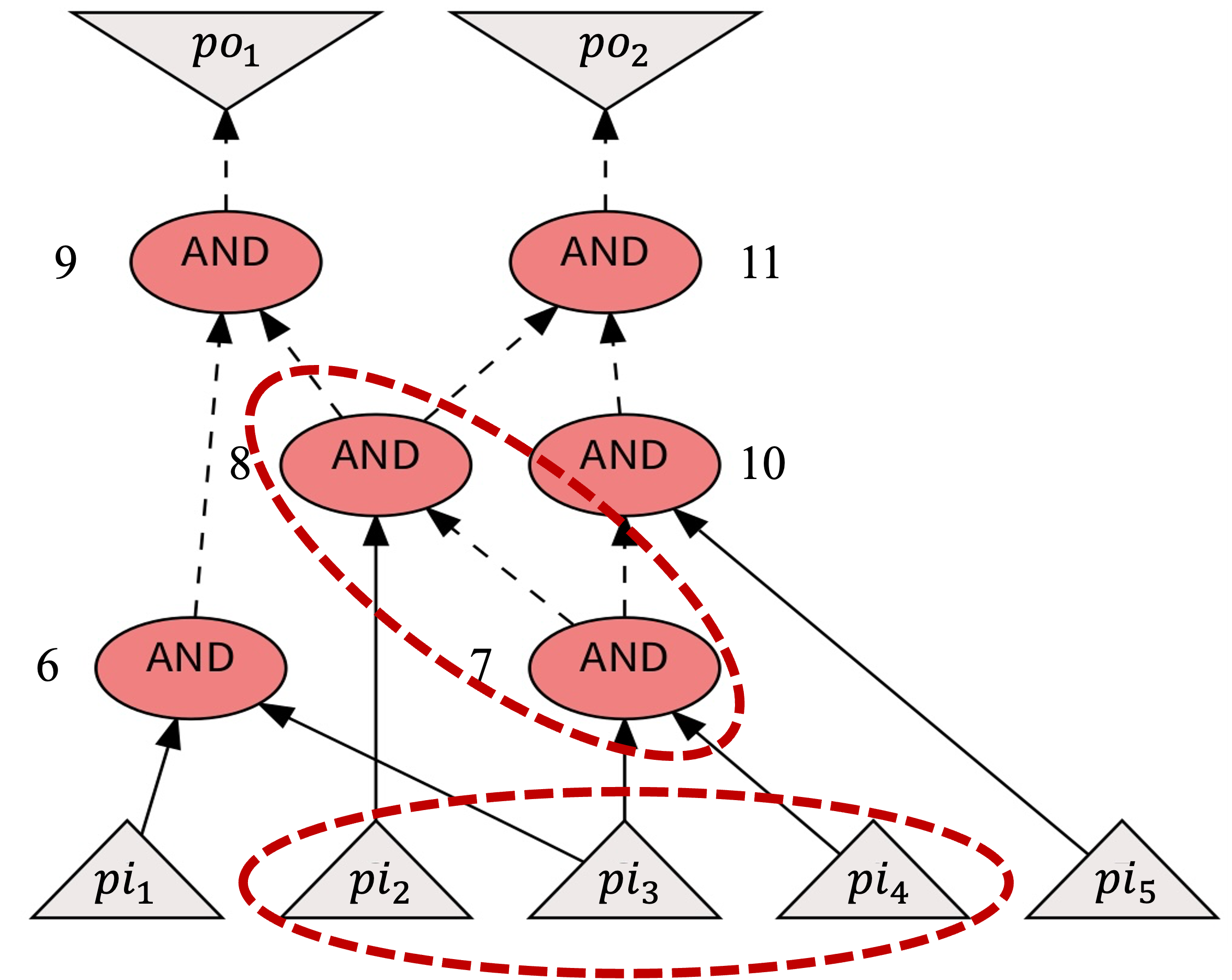}}
    \subfigure[pins \{$8$\} of cut \{$2,7$\}]{
    \label{fig:pin2}
    \includegraphics[width=0.4\linewidth]{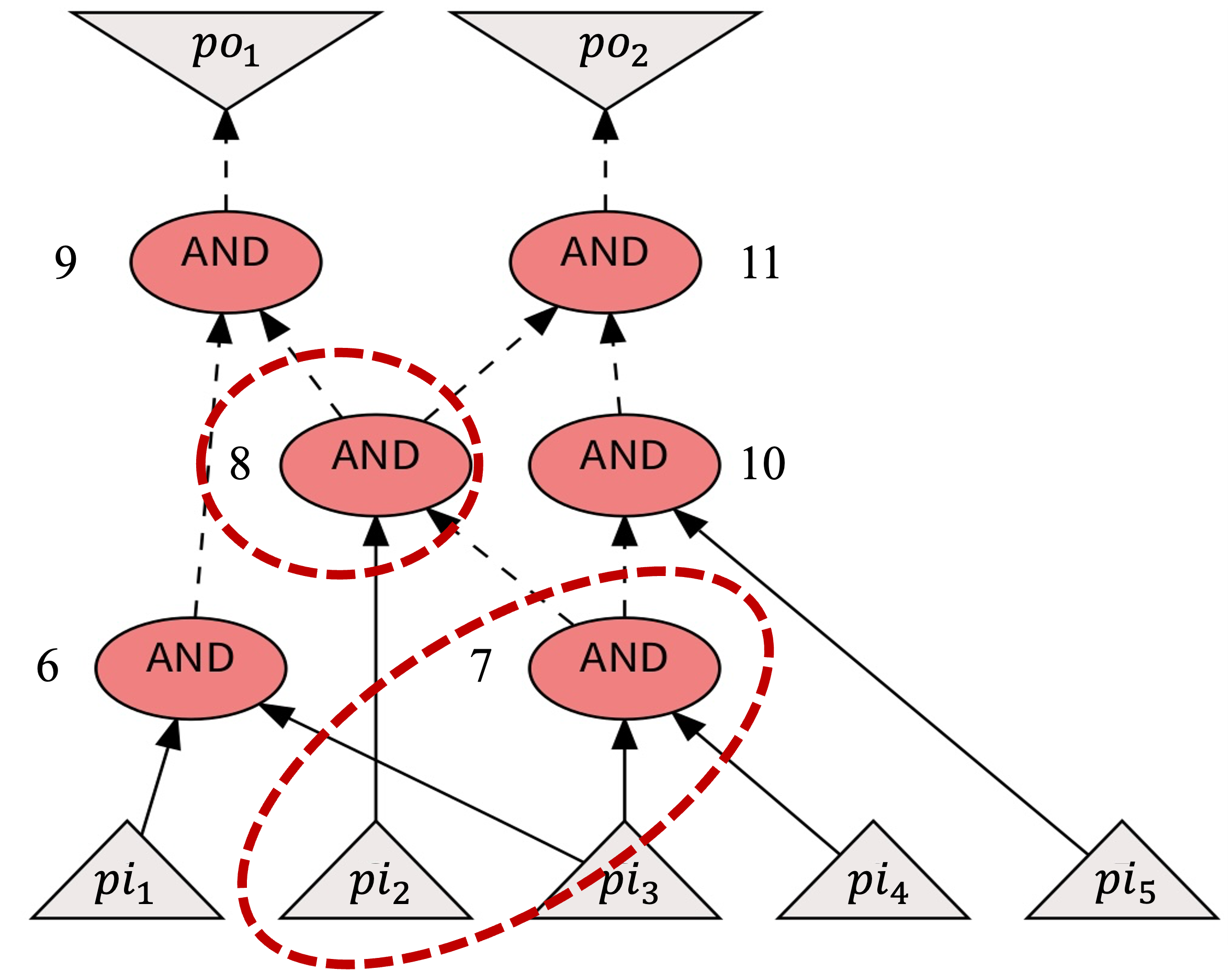}}
  \caption{An example of searching pins of node $8$. For node $8$, two different cuts are identified: \{$2,7$\} and \{$2,3,4$\}. Each cut has associated pins: the cut \{$2,7$\} is associated with the pin \{$8$\} and the cut \{$2,3,4$\} has pins \{$7,8$\}.}
  \label{fig:pin}
\end{figure}

For mapping preparation, the PigMAP-Mapper selects supergates that suit the corresponding cut function from the cell library and initializes parameters of each node. 
For each node in an AIG, PigMAP-Mapper enumerates all available cuts and supergates to search for their ``pins'', defined as the parent nodes of cuts' leaves in addition to cuts' roots.
The search for ``pins'' begins with the root of cut $c_i$ and recursively visits its children to determine if they are leaves, as shown in Algorithm~\ref{alg:search_pins}. 
The entire process is illustrated with the example shown in Figure~\ref{fig:pin}.
In this example, node $8$ has two different cuts, \{$2,7$\} and \{$2,3,4$\}.
For each group of ``pins'', we set a maximum search depth. 
If the depth is exceeded during the search, the process terminates immediately. 
The search is conducted by traversing from one level to the next (Lines 1-3).
Root (node $8$) is at first considered as a ``pin'' (Lines 4-5).
Once a child $3$ is confirmed as a leaf, its parent node $7$ is recognized as a ``pin'' and added to the ``pins'' set of $c_i$ (Lines 6-15).

\begin{algorithm}[t]
    \footnotesize
    \caption{Function \texttt{searchPin}.}
    \label{alg:search_pins}
    \begin{algorithmic}[1]
    \REQUIRE node $n$, cut $cut$, and level of search depth $l=0$
    \ENSURE pins $\mathbf{P}$
    \IF{$l$ > limit level of search depth $limit$}
        \RETURN $\mathbf{P}$;
    \ENDIF
    \STATE $l \gets l + 1$;
    \IF{size($cut$) == 0}
        \STATE $\mathbf{P} \leftarrow$ root of $cut$;
    \ENDIF
    \FORALL{fanin $child$ of $n$}
        \STATE $in\_cut \gets false$;
        \IF{$child$ in $cut$}
            \STATE $in\_cut \gets true$;
            \STATE break;
        \ENDIF
        \IF{$in\_cut == false$}
            \IF{$l > limit - 1$}
                \STATE $\mathbf{P} \leftarrow$ $child$;
                \RETURN $\mathbf{P}$;
            \ENDIF 
            \STATE $\mathbf{P} \leftarrow$ \texttt{searchPin}($child$, $cut$, $l$);
            \ELSE
                \RETURN $\mathbf{P}$;
        \ENDIF
    \ENDFOR
    \end{algorithmic}
\end{algorithm}

\begin{figure}[t]
  \centering  
  \subfigbottomskip=1pt 
  \subfigcapskip=-3pt 
    \subfigure[before mapping]{
    \label{fig:exam1}
    \includegraphics[width=0.4\linewidth]{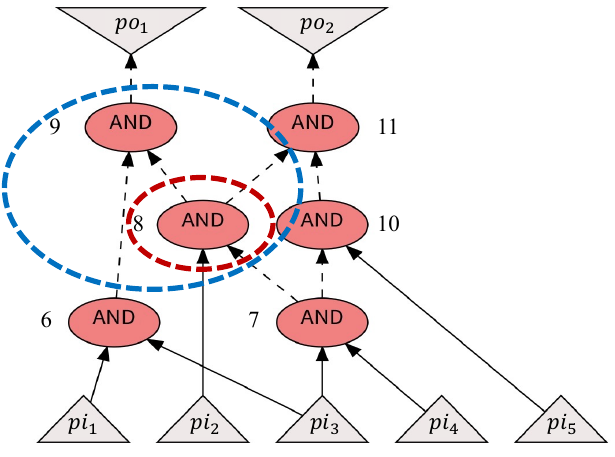}}
    \hspace{20mm}
    \subfigure[delay-driven mapping result]{
    \label{fig:exam2}
    \includegraphics[width=0.37\linewidth]{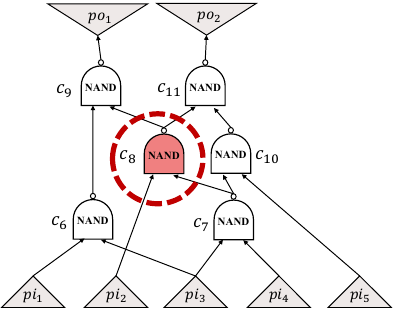}}
    \subfigure[performance-driven and power-driven mapping result]{
    \label{fig:exam3}
    \includegraphics[width=0.37\linewidth]{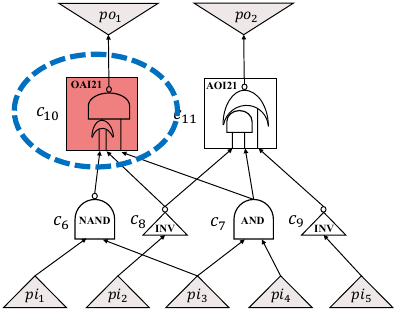}}
  \caption{An example of different strategies mapping. For delay-driven mapping, cuts \{$2,7$\} and \{$2,3,4$\} are analyzed to determine which results in the shorter delay. By choosing cut \{$2,7$\}, node $8$ is mapped as a 2-input NAND cell. For PigMAP-Performance mapping, we assess the wirelength from pins \{$2,7$\} to node $8$ for cut \{$2,7$\}, and from pins \{$7,8$\} to nodes \{$2,3,4$\} for cut \{$2,3,4$\}. However, since nodes $8$ and $9$ are far apart, we map node $8$ and $6$ together into a 3-input Or-And-Invert cell.}
  \label{fig:exam}
  \vspace{-2mm}
\end{figure}

After mapping preparation, we perform two optimization iterations of a wirelength-driven mapping.
The placement-aware mapping algorithm is detailed in Algorithm~\ref{alg:paflow}.
The mapping process is illustrated with the example shown in Figure~\ref{fig:exam}.
Considering that the logic function of each matched gate can be inverted by adding an inverter, we compute across both signal phases.
The algorithm initiates by setting four crucial metrics, critical path wirelength ($wl$), total wirelength ($total\_wl$), critical path delay ($delay$), and area ($area$), to $\infty$ (Lines 2-3).
The metrics $wl$ and $total\_wl$ represent new physically relative metrics known as ``virtual wirelength''.
Specifically, $wl$ denotes the maximum wirelength, which is formulated for performance-driven optimization. 
Similarly, $total\_wl$ represents the total wirelength, which is formulated for power-driven optimization. 
In this context, the performance-driven strategy focuses on reducing the maximum wirelength, whereas the power-driven strategy aims to minimize the total circuit wirelength.

\begin{owndefinition} {\bf Performance-driven cost}: Supposing the mapper is visiting node $n_i$, its cut is $c_i$, and its matched supergate is $g_i$.
The leaf of $n_i$ in order $l^\text{th}$ is $n_{i,l}$, and its cut and matched supergate are $c_{i,l}$ and $g_{i,l}$.
The coordinate of supergate $g_i$ is $g_i.x$, $g_i.y$.
The maximum wirelength of $g_i$ is the maximum wirelength among all paths set off from $g_i$ to every accessible PI, defined as
    \label{def2}
    \begin{equation}
        \label{eq:maxwl}
        \text{MW}(g_i)=\max_{0<i<l} \left(|g_i.x-g_{i,l}.x|+|g_i.y-g_{i,l}.y|\right)+\text{MW}(g_{i,l}),
    \end{equation}
    where $\text{MW}(g_{i,l})$ is the maximum wirelength of the best supergate matched for node $n_{i,l}$ in preceding iterations.
\end{owndefinition}

\begin{owndefinition} {\bf Power-driven cost}: The total wirelength from the visiting node to the PIs is computed by aggregating the discounted wirelength of all paths accessed by leaves.
The total wirelength is defined as
    \label{def3}
    \begin{small}
    \begin{equation}
    \label{eq:totalwl}
        \text{TW}(g_i)=\sum_{0<i<l} |g_i.x-g_{i,l}.x|+|g_i.y-g_{i,l}.y|+\text{TW}(g_{i,l})/\text{ERef},
    \end{equation}
    \end{small}%
    \noindent where the discount factor ERef accounts for the reference estimate of node $n_{i,l}$, helping to prevent redundant calculations of total wirelength for sub-circuits with multiple fanouts. 
\end{owndefinition}

\begin{algorithm}[t]
    \footnotesize
    \caption{Placement-aware Mapping Algorithm.}
    \label{alg:paflow}
    \begin{algorithmic}[1]
    \REQUIRE Optimization strategy $\mathbf{S}$, AIG $\mathbf{A}$, target node $n$
    \ENSURE Mapping solution $\mathbf{M}$
    \FORALL{none PI node $n$ of $\mathbf{A}$ in topological order}
        \FORALL{phase}
            \STATE $wl, total\_wl, delay, area \gets \infty$;
            \FORALL{$cut$ of $n$}
                \STATE $\mathbf{P}$= \texttt{computePosition}($cut$)
                \FORALL{$gate$ in \texttt{supergate}($cut$)}
                    \STATE $current\_delay \gets $ \texttt{compute\_delay}($n, gate, cut$);
                    \STATE $current\_area \gets$ \texttt{compute\_area}($n, gate, cut$);
                    \STATE $current\_wl \gets$ \texttt{compute\_wl}($n, gate, cut$);
                    \STATE $current\_total\_wl \gets$ \texttt{compute\_wl}($n, gate, cut$);
                    \IF{\texttt{setConstraint}($\mathbf{S}$)}
                        \STATE \textbf{continue};
                    \ENDIF
                    \IF{\texttt{comparePara}()}
                        \STATE $total\_wl \gets current\_total\_wl$;
                        \STATE $wl \gets current\_wl$;
                        \STATE $best\_gate \gets gate$;
                        \STATE $delay \gets current\_delay$;
                        \STATE $area \gets current\_area$;
                    \ENDIF
                \ENDFOR
            \ENDFOR
            \IF{\texttt{comparePara}()}
                \STATE \texttt{matchSel}($n$, $wl$, $total\_wl$, $best\_gate$, $delay$, $area$);
            \ENDIF
        \ENDFOR
    \ENDFOR
    \RETURN $\mathbf{M}$;
    \end{algorithmic}
\end{algorithm}

Afterwards, mapper computes approximate positions of all supergates by calculating the average position of pins associated with their corresponding cuts (Lines 4-5).
The resultant position-supergate tuples are stored to enable future wirelength computation.
Intending to find the best match for each node, the mapper enumerates all possible supergates and computes their metrics, including current delay ($current\_delay$), area ($current\_area$), and two distinct wirelength metrics: $current\_wl$ and $current\_total\_wl$ (Lines 6-10).
For node $8$ and cut \{$2,3,4$\}, $current\_wl$ of node $8$ represents the wirelength from pins \{$7,8$\} to nodes \{$2,3,4$\}, computed by equation~(\ref{eq:maxwl}).
For node $8$ and cut \{$2,7$\}, $current\_total\_wl$ of node $8$ represents the total wirelength from pins \{$8$\} to nodes \{$2,7$\}, computed by equation~(\ref{eq:totalwl}).
Since node $7$ serves as a fanin for nodes $8$ and $10$, the total wirelength computation for node $7$ is halved to distribute the power consumption.
Following metrics computation, we employ constraint-aided pruning to accelerate match selection,
removing matches that fail to meet the required arrival times or exceed maximum wirelength constraints. 
Under the PigMAP-Performance strategy, the mapper primarily focuses on the weighted sum of arrival time and maximum wirelength. 
Conversely, in the PigMAP-Power strategy, after discarding matches with excessive arrival times, total wirelength is prioritized, emphasizing reduction of power consumption (Lines 13-18).
In scenarios where gates exhibit similar timing performance due to delay and wirelength comparison, the algorithm advances to compare additional parameters such as area and cuts' size successively. 
Moreover, the power-driven mapping includes an additional global area mapping consideration to address potential area increases that could result from wirelength optimizations, recognizing that excessive area also contributes to higher power consumption.
Ultimately, a thorough comparison of these parameters identifies the best-performing supergate for each node, which is then saved for future reference. 
This saved supergate aids in subsequent wirelength computations when a higher-level node accesses it, ensuring optimized placement based on these strategic evaluations (Lines 19-20).
Finally, all optimal cuts of the entire network are determined using reverse topological sorting, with the mapping solution finalized (Line 21).

\section{EXPERIMENTAL RESULTS}
\label{sub:exp}

\begin{table*}[ht]
    \centering
    \fontsize{7}{7}\selectfont
    \caption{The Results reported by OpenROAD flow scripts~\cite{24_openroad_script} with ASAP 7nm technology.}
    \label{tab:exp3}\setlength\tabcolsep{0.7pt}
    \begin{tabularx}{\hsize}{@{}@{\extracolsep{\fill}}lcccrcccrcccrcccr@{}}
    \toprule
    \multirow{3}[0]{*}{Benchmark } & \multicolumn{4}{c}{Baseline~\cite{19PGMACTC_Ajayi_openroad}}  & \multicolumn{4}{c}{OpenPhySyn~\cite{20WOSET_Agiza_openphysyn}} & \multicolumn{4}{c}{PigMAP-Performance} & \multicolumn{4}{c}{PigMAP-Power} \\
    \cmidrule(lr){2-5} \cmidrule(lr){6-9} \cmidrule(lr){10-13} \cmidrule(lr){14-17}
    & Delay   & Area    & Power & \multicolumn{1}{c}{\multirow{2}[0]{*}{Time(s)}} & Delay   & Area    & Power & \multicolumn{1}{c}{\multirow{2}[0]{*}{Time(s)}} & Delay   & Area    & Power & \multicolumn{1}{c}{\multirow{2}[0]{*}{Time(s)}} & Delay   & Area    & Power & \multicolumn{1}{c}{\multirow{2}[0]{*}{Time(s)}} \\
    & ($ns$)  &  ($um^2$) & ($mW$)  &       & ($ns$)  &  ($um^2$) & ($mW$)  &       & ($ns$)  &  ($um^2$) & ($mW$)  &       & ($ns$)  &  ($um^2$) & ($mW$)  &  \\
    \midrule
    Adder & \multicolumn{1}{r}{6.069 } & \multicolumn{1}{r}{78.66 } & \multicolumn{1}{r}{0.02 } & 116.1  & \multicolumn{1}{r}{4.834 } & \multicolumn{1}{r}{87.48 } & \multicolumn{1}{r}{0.02 } & 141.1  & \multicolumn{1}{r}{4.883 } & \multicolumn{1}{r}{83.33 } & \multicolumn{1}{r}{0.02 } & 113.1  & \multicolumn{1}{r}{6.102 } & \multicolumn{1}{r}{78.80 } & \multicolumn{1}{r}{0.02 } & 184.1  \\
    Bar   & \multicolumn{1}{r}{1.009 } & \multicolumn{1}{r}{266.44 } & \multicolumn{1}{r}{0.84 } & 329.5  & \multicolumn{1}{r}{0.814 } & \multicolumn{1}{r}{265.36 } & \multicolumn{1}{r}{0.84 } & 345.5  & \multicolumn{1}{r}{0.791 } & \multicolumn{1}{r}{275.70 } & \multicolumn{1}{r}{0.90 } & 578.8  & \multicolumn{1}{r}{0.599 } & \multicolumn{1}{r}{268.99 } & \multicolumn{1}{r}{0.80 } & 589.8  \\
    Div   & \multicolumn{1}{r}{61.327 } & \multicolumn{1}{r}{5,514.20 } & \multicolumn{1}{r}{1.54 } & 3.553  & \multicolumn{4}{c}{N/A}        & \multicolumn{1}{r}{45.059 } & \multicolumn{1}{r}{5,813.00 } & \multicolumn{1}{r}{1.80 } & 4,399.4  & \multicolumn{1}{r}{64.989 } & \multicolumn{1}{r}{5,150.50 } & \multicolumn{1}{r}{1.26 } & 3,938.1  \\
    Hyp   & \multicolumn{1}{r}{362.078 } & \multicolumn{1}{r}{13,431.56 } & \multicolumn{1}{r}{13.14 } & 7,933.5  & \multicolumn{4}{c}{N/A}        & \multicolumn{1}{r}{357.517 } & \multicolumn{1}{r}{17,103.80 } & \multicolumn{1}{r}{15.38 } & 7,814.5  & \multicolumn{1}{r}{363.869 } & \multicolumn{1}{r}{16,640.60 } & \multicolumn{1}{r}{10.40 } & 7,756.2  \\
    Log2  & \multicolumn{1}{r}{12.417 } & \multicolumn{1}{r}{1,516.47 } & \multicolumn{1}{r}{0.65 } & 942.8  & \multicolumn{1}{r}{10.958 } & \multicolumn{1}{r}{1,509.50 } & \multicolumn{1}{r}{1.00 } & 1,240.8  & \multicolumn{1}{r}{7.202 } & \multicolumn{1}{r}{1,854.44 } & \multicolumn{1}{r}{0.71 } & 1,488.4  & \multicolumn{1}{r}{10.521 } & \multicolumn{1}{r}{1,642.15 } & \multicolumn{1}{r}{0.68 } & 1,663.9  \\
    Max   & \multicolumn{1}{r}{3.558 } & \multicolumn{1}{r}{225.01 } & \multicolumn{1}{r}{0.19 } & 339.6  & \multicolumn{1}{r}{3.351 } & \multicolumn{1}{r}{225.23 } & \multicolumn{1}{r}{0.18 } & 339.6  & \multicolumn{1}{r}{3.381 } & \multicolumn{1}{r}{233.75 } & \multicolumn{1}{r}{0.21 } & 526.8  & \multicolumn{1}{r}{\cellcolor[rgb]{ .867,  .922,  .969}\textbf{3.463 }} & \multicolumn{1}{r}{\cellcolor[rgb]{ .867,  .922,  .969}\textbf{215.26 }} & \multicolumn{1}{r}{\cellcolor[rgb]{ .867,  .922,  .969}\textbf{0.17 }} & \cellcolor[rgb]{ .867,  .922,  .969}\textbf{550.8 } \\
    Multiplier & \multicolumn{1}{r}{6.400 } & \multicolumn{1}{r}{1,187.95 } & \multicolumn{1}{r}{1.46 } & 883.0  & \multicolumn{4}{c}{N/A}        & \multicolumn{1}{r}{4.619 } & \multicolumn{1}{r}{1,511.97 } & \multicolumn{1}{r}{1.61 } & 633.0  & \multicolumn{1}{r}{6.053 } & \multicolumn{1}{r}{1,389.86 } & \multicolumn{1}{r}{1.42 } & 1,526.2  \\
    Sin   & \multicolumn{1}{r}{4.328 } & \multicolumn{1}{r}{317.38 } & \multicolumn{1}{r}{0.13 } & 553.9  & \multicolumn{4}{c}{N/A}        & \multicolumn{1}{r}{3.743 } & \multicolumn{1}{r}{356.62 } & \multicolumn{1}{r}{0.17 } & 530.1  & \multicolumn{1}{r}{4.418 } & \multicolumn{1}{r}{336.51 } & \multicolumn{1}{r}{0.12 } & 733.2  \\
    Sqrt  & \multicolumn{1}{r}{105.363 } & \multicolumn{1}{r}{2,170.41 } & \multicolumn{1}{r}{3.24 } & 1,132.6  & \multicolumn{1}{r}{98.392 } & \multicolumn{1}{r}{2,184.75 } & \multicolumn{1}{r}{4.24 } & 2,184.60  & \multicolumn{1}{r}{\cellcolor[rgb]{ .867,  .922,  .969}\textbf{89.621 }} & \multicolumn{1}{r}{\cellcolor[rgb]{ .867,  .922,  .969}\textbf{1,940.95 }} & \multicolumn{1}{r}{\cellcolor[rgb]{ .867,  .922,  .969}\textbf{3.16 }} & \cellcolor[rgb]{ .867,  .922,  .969}\textbf{1,478.4 } & \multicolumn{1}{r}{\cellcolor[rgb]{ .867,  .922,  .969}\textbf{101.400 }} & \multicolumn{1}{r}{\cellcolor[rgb]{ .867,  .922,  .969}\textbf{1,600.52 }} & \multicolumn{1}{r}{\cellcolor[rgb]{ .867,  .922,  .969}\textbf{2.88 }} & \cellcolor[rgb]{ .867,  .922,  .969}\textbf{1,128.4 } \\
    Square & \multicolumn{1}{r}{6.703 } & \multicolumn{1}{r}{1,016.36 } & \multicolumn{1}{r}{0.75 } & 849.3  & \multicolumn{1}{r}{6.658 } & \multicolumn{1}{r}{1,017.22 } & \multicolumn{1}{r}{0.75 } & 782.31  & \multicolumn{1}{r}{4.620 } & \multicolumn{1}{r}{1,419.43 } & \multicolumn{1}{r}{0.82 } & 1,218.8  & \multicolumn{1}{r}{5.899 } & \multicolumn{1}{r}{1,243.71 } & \multicolumn{1}{r}{0.74 } & 1,264.7  \\
    Arbiter & \multicolumn{1}{r}{1.134 } & \multicolumn{1}{r}{911.24 } & \multicolumn{1}{r}{1.32 } & 564.8  & \multicolumn{1}{r}{1.109 } & \multicolumn{1}{r}{909.08 } & \multicolumn{1}{r}{1.32 } & 886.81  & \multicolumn{1}{r}{\cellcolor[rgb]{ .867,  .922,  .969}\textbf{1.066 }} & \multicolumn{1}{r}{\cellcolor[rgb]{ .867,  .922,  .969}\textbf{796.00 }} & \multicolumn{1}{r}{\cellcolor[rgb]{ .867,  .922,  .969}\textbf{0.95 }} & \cellcolor[rgb]{ .867,  .922,  .969}\textbf{297.2 } & \multicolumn{1}{r}{\cellcolor[rgb]{ .867,  .922,  .969}\textbf{1.101 }} & \multicolumn{1}{r}{\cellcolor[rgb]{ .867,  .922,  .969}\textbf{804.13 }} & \multicolumn{1}{r}{\cellcolor[rgb]{ .867,  .922,  .969}\textbf{0.93 }} & \cellcolor[rgb]{ .867,  .922,  .969}\textbf{1,108.5 } \\
    Cavlc & \multicolumn{1}{r}{0.310 } & \multicolumn{1}{r}{40.81 } & \multicolumn{1}{r}{0.12 } & 226.1  & \multicolumn{1}{r}{0.311 } & \multicolumn{1}{r}{41.14 } & \multicolumn{1}{r}{0.13 } & 304.09  & \multicolumn{1}{r}{0.304 } & \multicolumn{1}{r}{42.34 } & \multicolumn{1}{r}{0.13 } & 361.1  & \multicolumn{1}{r}{0.304 } & \multicolumn{1}{r}{42.34 } & \multicolumn{1}{r}{0.13 } & 400.1  \\
    Ctrl  & \multicolumn{1}{r}{0.165 } & \multicolumn{1}{r}{13.38 } & \multicolumn{1}{r}{0.10 } & 88.0  & \multicolumn{1}{r}{0.160 } & \multicolumn{1}{r}{12.54 } & \multicolumn{1}{r}{0.08 } & 159.02  & \multicolumn{1}{r}{\cellcolor[rgb]{ .867,  .922,  .969}\textbf{0.145 }} & \multicolumn{1}{r}{\cellcolor[rgb]{ .867,  .922,  .969}\textbf{11.84 }} & \multicolumn{1}{r}{\cellcolor[rgb]{ .867,  .922,  .969}\textbf{0.07 }} & \cellcolor[rgb]{ .867,  .922,  .969}\textbf{209.0 } & \multicolumn{1}{r}{\cellcolor[rgb]{ .867,  .922,  .969}\textbf{0.159 }} & \multicolumn{1}{r}{\cellcolor[rgb]{ .867,  .922,  .969}\textbf{12.64 }} & \multicolumn{1}{r}{\cellcolor[rgb]{ .867,  .922,  .969}\textbf{0.08 }} & \cellcolor[rgb]{ .867,  .922,  .969}\textbf{213.0 } \\
    Dec   & \multicolumn{1}{r}{0.184 } & \multicolumn{1}{r}{53.13 } & \multicolumn{1}{r}{0.10 } & 150.1  & \multicolumn{1}{r}{0.184 } & \multicolumn{1}{r}{53.74 } & \multicolumn{1}{r}{0.10 } & 186.06  & \multicolumn{1}{r}{0.187 } & \multicolumn{1}{r}{60.70 } & \multicolumn{1}{r}{0.14 } & 185.1  & \multicolumn{1}{r}{0.218 } & \multicolumn{1}{r}{60.01 } & \multicolumn{1}{r}{0.10 } & 303.1  \\
    I2c   & \multicolumn{1}{r}{0.259 } & \multicolumn{1}{r}{110.33 } & \multicolumn{1}{r}{0.48 } & 222.1  & \multicolumn{1}{r}{0.254 } & \multicolumn{1}{r}{108.68 } & \multicolumn{1}{r}{0.46 } & 323.15  & \multicolumn{1}{r}{0.245 } & \multicolumn{1}{r}{111.00 } & \multicolumn{1}{r}{0.48 } & 304.2  & \multicolumn{1}{r}{0.245 } & \multicolumn{1}{r}{111.00 } & \multicolumn{1}{r}{0.48 } & 401.2  \\
    Int2float & \multicolumn{1}{r}{0.186 } & \multicolumn{1}{r}{16.53 } & \multicolumn{1}{r}{0.20 } & 151.0  & \multicolumn{1}{r}{0.191 } & \multicolumn{1}{r}{16.87 } & \multicolumn{1}{r}{0.20 } & 192.03  & \multicolumn{1}{r}{0.172 } & \multicolumn{1}{r}{17.35 } & \multicolumn{1}{r}{0.20 } & 230.0  & \multicolumn{1}{r}{0.188 } & \multicolumn{1}{r}{17.74 } & \multicolumn{1}{r}{0.19 } & 271.0  \\
    Mem\_ctrl & \multicolumn{1}{r}{2.528 } & \multicolumn{1}{r}{2,997.41 } & \multicolumn{1}{r}{1.64 } & 1,498.3  & \multicolumn{1}{r}{2.535 } & \multicolumn{1}{r}{3,000.87 } & \multicolumn{1}{r}{1.64 } & 1,759.31  & \multicolumn{1}{r}{2.502 } & \multicolumn{1}{r}{3,244.81 } & \multicolumn{1}{r}{1.75 } & 1,262.6  & \multicolumn{1}{r}{2.608 } & \multicolumn{1}{r}{3,158.92 } & \multicolumn{1}{r}{1.54 } & 2,382.7  \\
    Priority & \multicolumn{1}{r}{2.540 } & \multicolumn{1}{r}{75.15 } & \multicolumn{1}{r}{0.03 } & 164.1  & \multicolumn{1}{r}{2.526 } & \multicolumn{1}{r}{76.12 } & \multicolumn{1}{r}{0.03 } & 204.15  & \multicolumn{1}{r}{2.227 } & \multicolumn{1}{r}{87.03 } & \multicolumn{1}{r}{0.04 } & 242.2  & \multicolumn{1}{r}{2.832 } & \multicolumn{1}{r}{81.22 } & \multicolumn{1}{r}{0.02 } & 310.2  \\
    Router & \multicolumn{1}{r}{0.583 } & \multicolumn{1}{r}{23.01 } & \multicolumn{1}{r}{0.03 } & 135.0  & \multicolumn{1}{r}{0.586 } & \multicolumn{1}{r}{22.92 } & \multicolumn{1}{r}{0.03 } & 159.04  & \multicolumn{1}{r}{0.611 } & \multicolumn{1}{r}{22.73 } & \multicolumn{1}{r}{0.03 } & 204.1  & \multicolumn{1}{r}{0.628 } & \multicolumn{1}{r}{21.93 } & \multicolumn{1}{r}{0.03 } & 250.1  \\
    Voter & \multicolumn{1}{r}{1.644 } & \multicolumn{1}{r}{978.60 } & \multicolumn{1}{r}{2.07 } & 664.7  & \multicolumn{1}{r}{1.610 } & \multicolumn{1}{r}{974.02 } & \multicolumn{1}{r}{2.09 } & 603.71  & \multicolumn{1}{r}{1.341 } & \multicolumn{1}{r}{1,089.46 } & \multicolumn{1}{r}{2.13 } & 945.2  & \multicolumn{1}{r}{1.625 } & \multicolumn{1}{r}{988.98 } & \multicolumn{1}{r}{1.90 } & 964.1  \\
    \midrule
    \textbf{Geo.} & \multicolumn{1}{r}{2.682} & \multicolumn{1}{r}{317.41 } & \multicolumn{1}{r}{0.40 } & 462.77  & \multicolumn{1}{r}{$1.423^*$ } & \multicolumn{1}{r}{$194.32^*$ } & \multicolumn{1}{r}{$0.30^*$ } & $408.42^*$  & \multicolumn{1}{r}{2.308 } & \multicolumn{1}{r}{342.02 } & \multicolumn{1}{r}{0.42 } & 579.6  & \multicolumn{1}{r}{2.602 } & \multicolumn{1}{r}{324.66 } & \multicolumn{1}{r}{0.36 } & 743.9  \\
    \textbf{Ratio.} & \multicolumn{1}{r}{1.0 } & \multicolumn{1}{r}{1.0 } & \multicolumn{1}{r}{1.0 } & 1.0   & \multicolumn{1}{r}{\textbf{$0.95^*$ }} & \multicolumn{1}{r}{\textbf{$1.00^*$ }} & \multicolumn{1}{r}{\textbf{$1.03^*$ }} & $1.41^*$  & \multicolumn{1}{r}{\textbf{0.86 }} & \multicolumn{1}{r}{\textbf{1.08 }} & \multicolumn{1}{r}{\textbf{1.06 }} & 1.25  & \multicolumn{1}{r}{\textbf{0.97 }} & \multicolumn{1}{r}{\textbf{1.02 }} & \multicolumn{1}{r}{\textbf{0.91 }} & 1.61  \\
    \textbf{Average} & \multicolumn{3}{c}{1.0 } &       & \multicolumn{3}{c}{\textbf{$0.982^*$ }} &       & \multicolumn{3}{c}{\textbf{0.980 }} &       & \multicolumn{3}{c}{\textbf{0.902 }} &  \\
    \bottomrule
    \end{tabularx}
    \begin{tablenotes}
        \footnotesize
        \item[1] {\bf Geo.}: The geometric mean of the above dataset, {\bf Ratio.}: Average geometric mean improvement (new/old)
        \item[2] {\bf Average}: Average Delay improvement per unit area and per unit power consumption (new/old).
        \item[3] {\bf *}: Data with this superscript derived only from successful processing (excluding N/A data).
    \end{tablenotes}
    \vspace{-4mm}
\end{table*}

We implement our PigMAP framework in the C++ programming language and the logic synthesis library mockturtle~\cite{24_EPFL_mockturtle} for cut enumeration and technology mapping. 
We run all experiments on a Linux server with an Intel Xeon Gold 6226R 2.9GHz CPU and 128GB memory.
To show the effectiveness and efficiency of our method, the following four scenarios are evaluated and compared.
\begin{itemize}[leftmargin=*,itemsep=3pt, topsep=2pt]
    \item {\em Baseline}: 
    We first perform the logic synthesis using the open-source Yosys suite~\cite{16_Wolf_yosys}, and then push synthesized netlist into back-end tools for place-and-route. This is the default flow in the OpenROAD flow scripts~\cite{24_openroad_script}.
    \item {\em OpenPhySyn}: We perform the conventional physical synthesis flow using OpenPhySyn~\cite{20WOSET_Agiza_openphysyn}.
     We follow the exact steps in~\cite{20WOSET_Agiza_openphysyn} to use OpenPhySyn for physical synthesis, as shown in Figure~\ref{fig:flowps}(a).
    \item {\em PigMAP-Performance}/{\em PigMAP-Power}: We first perform the physically aware logic synthesis using PigMAP with the PigMAP-Performance/PigMAP-Power mapping enabled, and then push the synthesized netlist into back-end tools for place-and-route.
\end{itemize}

The evaluation flow for {\em PigMAP-Performance} and {\em PigMAP-Power} is demonstrated in Figure~\ref{fig:flowps}(b).
The physical design is conducted with the latest OpenROAD flow~\cite{19PGMACTC_Ajayi_openroad} in 7nm open-source technology node (ASAP7~\cite{16MJ_Clark_asap7}). 
PPA results are collected from the routed design.
For each design evaluated, we ensure that the core area is consistent while following the default parameters set by OpenROAD, typically aiming for 50\% utilization.
We evaluate different scenarios using the EPFL combinational arithmetic benchmark suite~\cite{23_epfl}.
Given that our benchmarks consist entirely of combinational logics, OpenROAD by default creates a virtual clock. 
We carefully assign tight clock period for each benchmark and ensure that the final slack is zero, thus allowing us to accurately measure the effectiveness of PigMAP in optimizing performance and power.

\vspace{-2mm}
\subsection{Experimental Results and Analysis}

For each PigMAP scheme, our experiments are divided into two main phases. 
The first phase analyzes both the synthesis and routing results to demonstrate how the virtual wirelength constraint assists the mapper in achieving a PD-friendly netlist.
The results are visually represented in Figure~\ref{fig:result}, which shows the trend of various metrics including mapped netlist area (Post-synthesis area), mapped netlist delay (Post-synthesis delay), estimated critical path wirelength (Virtual wirelength), estimated total wirelength (Virtual total wirelength), post-routing total wirelength (Total WL), total number of vias (Via count), and Post-routing critical path delay (Post-routing delay).
The ordinate represents the average geometric mean improvement of delay (PigMAP/Baseline).

The second phase includes end-to-end experiments focusing on evaluating the efficiency and effectiveness across the entire flow process.
The results are shown in Table~\ref{tab:exp3}, providing a detailed comparison that includes the final delay (Delay), standard cell area (Area), power consumption (Power), and total runtime (Time) for each method. 
To ensure a fair comparison, we assess the improvement in critical path delay relative to both unit area and unit power consumption ({\bf Average}). 
This comprehensive listing allows for a thorough evaluation of the performance across different approaches.
The experimental results encompass several large benchmarks, which exhibit excessive delays.
However, the results align with~\cite{20_shahsavani_tdp}, confirming the reliability of our experiments.

\subsubsection{PigMAP-Performance. }

\begin{figure*}[t]
  \centering  
  \includegraphics[width=0.9\linewidth]{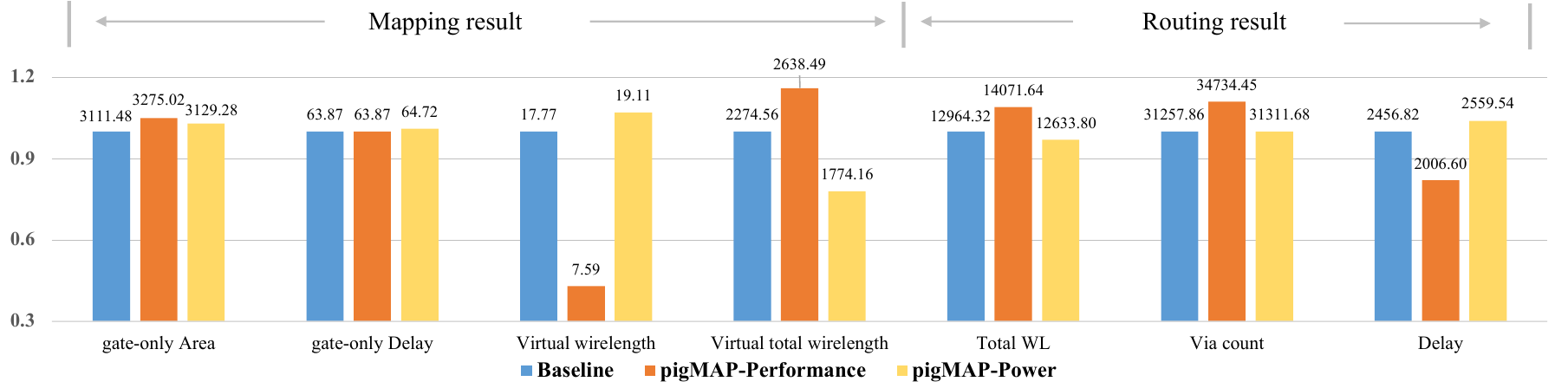}
  \caption{The mapping and routing results of PigMAP.}
  \label{fig:result}
  \vspace{-2mm}
\end{figure*}

During the mapping phase, as shown in Figure~\ref{fig:result}, PigMAP-Performance achieves up to a 57\% improvement in virtual wirelength at the cost of a 5\% increase in area, while maintaining a consistent delay. 
This focus on optimizing the critical path, however, does not extend to non-critical paths, resulting in an 18\% increase in total virtual wirelength. 
Transitioning to the routing phase, PigMAP-Performance demonstrates a significant correlation between the improvement in virtual wirelength and a 14\% improvement in delay. 
Additionally, the 5\% increase in area correlates with an 11\% rise in via count.

Overall, PigMAP-Performance achieves a notable 14\% reduction in delay, with only an 8\% increase in area and a 6\% rise in power consumption compared to the baseline, as shown in Table~\ref{tab:exp3}. 
This mode shows comprehensive improvements in PPA metrics for 3 out of 20 benchmarks, indicating substantial progress. 
Conversely, OpenPhySyn encounters issues in 4 out of 20 benchmarks, failing to produce results due to errors during physically aware synthesis.
In the 16 benchmarks where it completed successfully, PigMAP-Performance manages a 9\% reduction in delay but also experiences a 3\% increase in power consumption compared to OpenPhySyn.
When considering average delay improvement relative to the baseline, both PigMAP-Performance and OpenPhySyn exhibit similar levels of enhancement, each improving by about 2\%.
These results underscore that PigMAP-Performance can effectively enhance final performance by minimizing the wirelength of critical paths during the mapping phase. 
This optimization approach provides a targeted solution that improves overall design performance.

\subsubsection{PigMAP-Power. }

Similarly, during the mapping phase, PigMAP-Power slightly increases the delay by 1\% but achieves a significant 22\% improvement in virtual total wirelength by focusing on optimizing the global wirelength rather than prioritizing the critical path wirelength. 
This broad focus leads to some optimization, resulting in only an 8\% increase in the virtual wirelength of the critical path.
Transitioning to the routing phase, PigMAP-Power demonstrates how improvements in virtual total wirelength correlate to a 3\% improvement in total WL. 

Overall, PigMAP-Power realizes a 3\% reduction in delay and a notable 9\% decrease in power consumption, with just a 3\% increase in area compared to the baseline. 
This mode also shows comprehensive PPA improvements for 4 out of 20 benchmarks.
Compared to OpenPhySyn, PigMAP-Power achieves a notable 12\% reduction in power consumption, with only a 2\% increase in delay and a 2\% increase in area. 
In terms of average delay improvement, PigMAP-Power demonstrates a more substantial performance improvement of 10\%.
These results highlight that PigMAP-Power can effectively minimize global wirelength during the mapping phase while also optimizing critical path delay, leading to significant improvements in final power consumption and performance.
This optimization approach provides a comprehensive solution, enhancing overall circuit efficiency.

Referring back to Table~\ref{tab:exp3}, we find that the virtual wirelength during the mapping stage can influence post-routing delay, thus enhancing final performance. 
Similarly, the virtual total wirelength during the mapping phase can impact the post-routing total WL, consequently reducing final power consumption.
These results clearly demonstrate how our wirelength-driven mapping algorithm significantly affects routing outcomes and decisively influences the final PPA.
Additionally, PigMAP-Performance exhibits a total runtime that is 1.25$\times$ slower, while PigMAP-Power shows a total runtime that is 1.61$\times$ slower. 
Given the varying loads in each test, fluctuations in time of up to 50\% are considered normal. 

\begin{figure}[t]
  \centering  
  \subfigbottomskip=2pt 
  \subfigcapskip=-5pt 
  \subfigure[Critical path of OpenROAD flow~\cite{24_openroad_script}]{
    \label{fig:baseline}
    \includegraphics[width=0.39\linewidth]{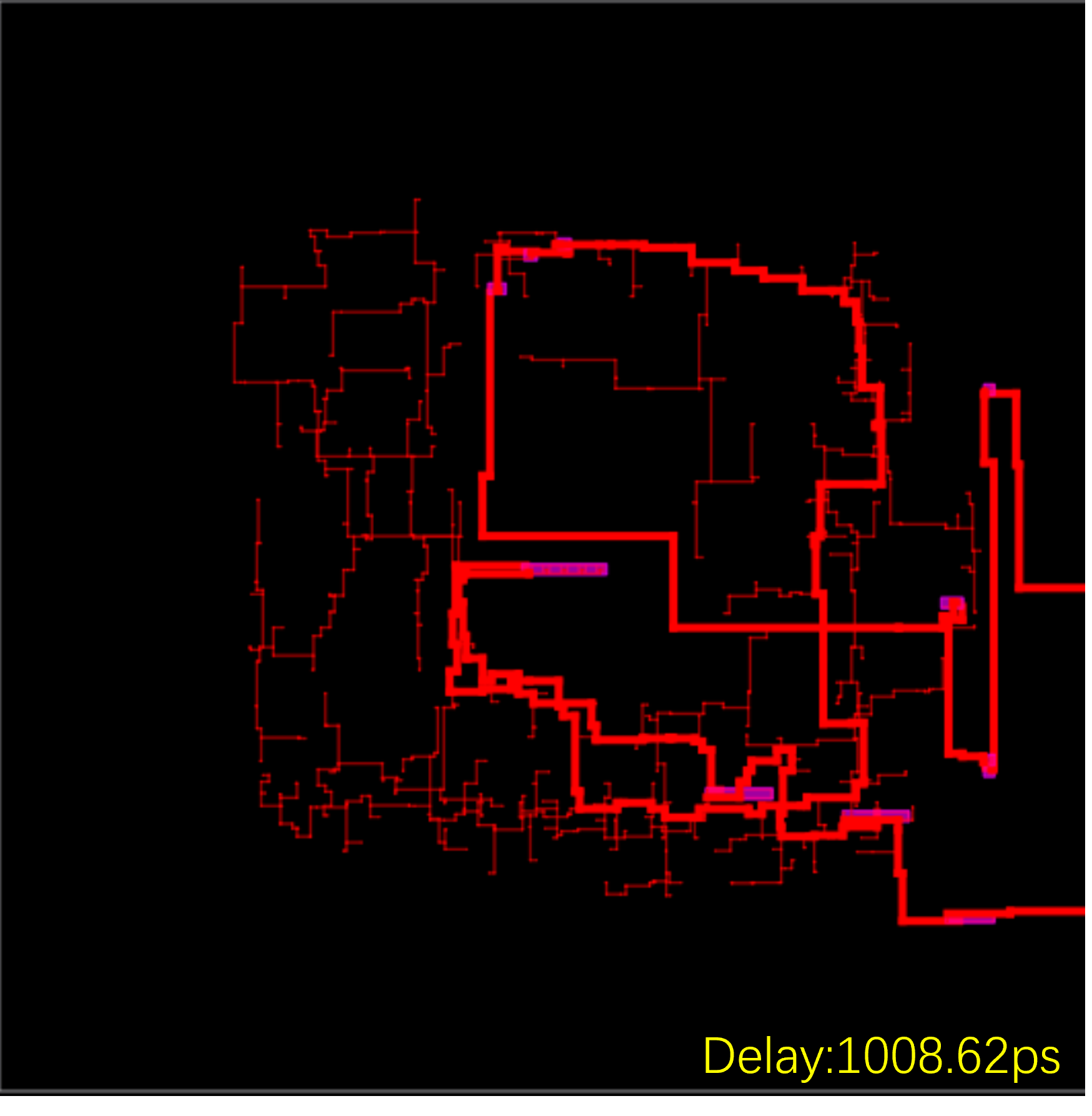}
  }
  \hspace{5mm}
    \subfigure[Critical path of OpenPhySyn~\cite{20WOSET_Agiza_openphysyn}]{
    \label{fig:openphysyn}
    \includegraphics[width=0.39\linewidth]{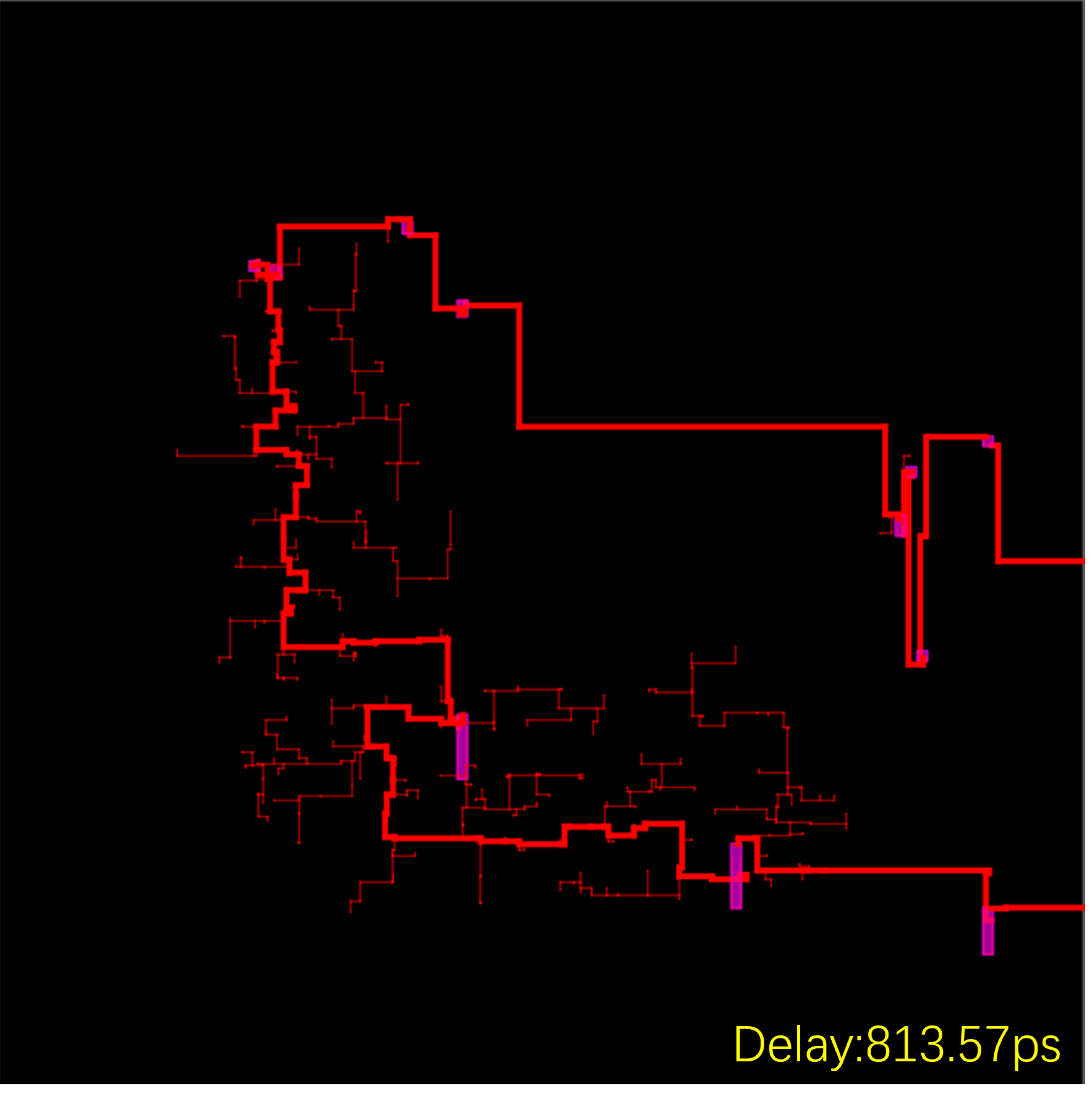}
  }
  \subfigure[Critical path of PigMAP-Performance]{
    \label{fig:performance-driven PigMAP}
    \includegraphics[width=0.39\linewidth]{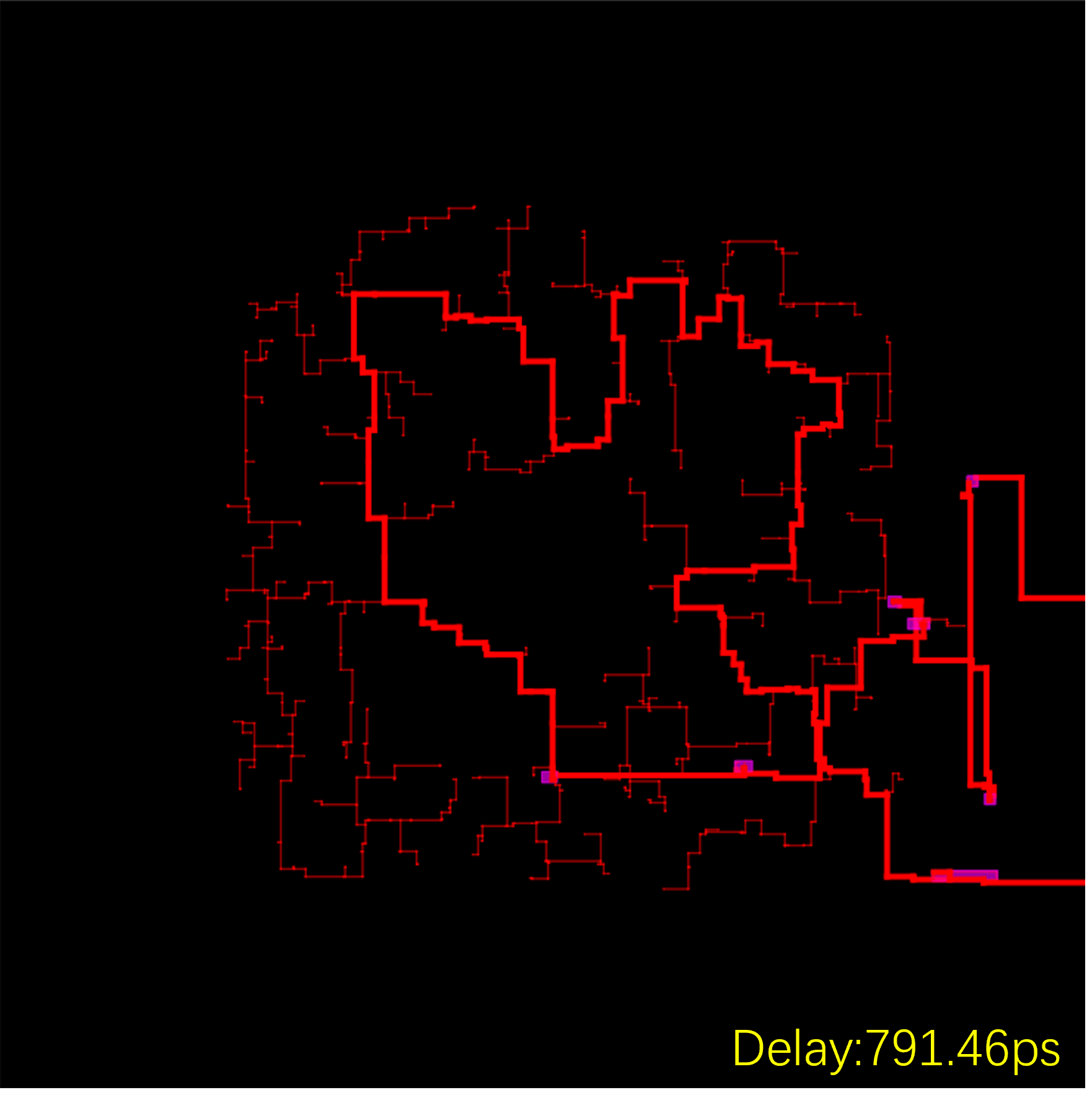}
  }
  \hspace{5mm}
  \subfigure[PigMAP-Power path that shares the same I/O as critical path of OpenROAD flow]{
    \label{fig:power-driven PigMAP}
    \includegraphics[width=0.39\linewidth]{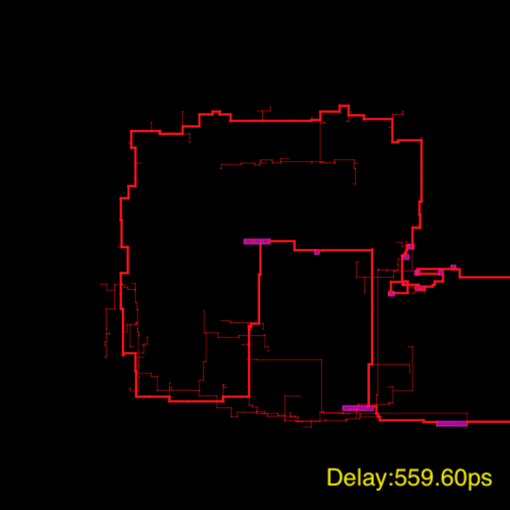}
  }
  \caption{The final layout of different methods captured from EPFL benchmark Bar, critical path is indicated with red lines.}
  \label{fig:layout}
  \vspace{-4mm}
\end{figure}

\subsection{Case Study}
\label{subsec:cs}

In this subsection, we employ the {\em Bar} benchmark to demonstrate how our approach effectively optimizes performance and power.
The conventional physical synthesis method (OpenPhySyn) effectively reduces the delay of critical paths and enhances performance, as depicted in Figure~\ref{fig:baseline} and Figure~\ref{fig:openphysyn}.
As illustrated in Figure~\ref{fig:performance-driven PigMAP}, the critical path resulting from the PigMAP-Performance method shows fewer back-and-forth routing paths. 
This streamlined routing effectively reduces wirelength consumption, leading to significant improvements in timing performance.
For PigMAP-Power, as demonstrated in Figure~\ref{fig:power-driven PigMAP}, we select a path with the same I/O configuration of the OpenROAD flow critical path for comparison. 
On a non-critical path (highlighted in light red), PigMAP-Power reduces the interconnects between instances, effectively cutting the total wirelength. 
This reduction in interconnects substantially decreases power consumption, underscoring the energy-saving effectiveness of our approach.
The routing paths chosen by PigMAP exhibit fewer detours and an overall reduction in wirelength, indicating strategic differences from conventional methods. 
While conventional methods typically focus on addressing local structural issues, our implementation of global operations fundamentally alters the routing structure. 
This change significantly improves performance outcomes, showcasing the effectiveness of our wirelength-driven strategy in enhancing critical path delay and power efficiency in circuit design.

\section{CONCLUSION}
\label{sec:con}

The introduction of PigMAP in this paper addresses the challenge of incorporating physical awareness into the logic synthesis mapping process. 
By utilizing wirelength constraints to direct the mapping search, PigMAP not only enhances the connection between logic synthesis and the final routed design but also secures notable improvements in both critical path delay and power consumption.
The framework is equipped with two operational schemes designed to meet distinct optimization goals: PigMAP-Performance and PigMAP-Power.
The PigMAP-Performance is geared towards minimizing critical path delay, achieving a substantial 14\% reduction, though this comes with a slight increase in power consumption. 
On the other hand, the PigMAP-Power is designed specifically to cut power consumption, achieving a significant decrease of 9\%, while also delivering a modest 3\% improvement in delay.

\begin{acks}
This work was supported by the National Natural Science Foundation of China (NSFC) Research Projects under Grant 92373207.
\end{acks}

\clearpage
\balance
\bibliographystyle{ACM-Reference-Format}
\bibliography{Top_sim, ref}

\end{sloppypar}
\end{document}